\documentclass[12pt, centerh1]{article}
\usepackage{amssymb,amsmath,colonequals,color,multirow,natbib,longtable}
\usepackage{amsmath, amsthm, amsfonts, amssymb, amscd, bm}

 \usepackage{graphicx}

\usepackage{marvosym}
\usepackage{algorithm}
\usepackage{verbatim}
\usepackage{bm,relsize}
\usepackage{geometry}
\usepackage{lscape}
\usepackage{mathtools}
\usepackage{caption}
\usepackage{theorem}
\usepackage{placeins}
\usepackage{enumerate}
\usepackage{float}
\usepackage{caption}
\usepackage{subcaption}
\usepackage[draft]{hyperref}
\usepackage{tablefootnote}

\begin{document}

\title{Model Based Clustering of High-Dimensional Binary Data}

\author{Yang Tang \qquad Ryan~P.~Browne\thanks{Department of Mathematics \&Statistics, University of Guelph, Guelph, Ontario, N1G 2W1, Canada. E-mail: rbrowne@uoguelph.ca.} \qquad Paul D.\ McNicholas
}
\date{}

\maketitle

\begin{abstract}
We propose a mixture of latent trait models with common slope parameters (MCLT) for model-based clustering of high-dimensional binary data, a data type for which few established methods exist. Recent work on clustering of binary data, based on a $d$-dimensional Gaussian latent variable, is extended by incorporating common factor analyzers. Accordingly, our approach facilitates a low-dimensional visual representation of the clusters. We extend the model further by the incorporation of random block effects. The dependencies in each block are taken into account through block-specific parameters that are considered to be random variables. A variational approximation to the likelihood is exploited to derive a fast algorithm for determining the model parameters. Our approach is demonstrated on real and simulated data. 
\end{abstract}


\section{Introduction}\label{sec:intro}
Binary manifest variables are extremely common in behavioural and social sciences research, e.g., individuals may be classified according to whether they take holidays abroad or whether they are satisfied with their lives. In such circumstances, they can be recorded as agreeing or disagreeing with some proposition, or as being capable of doing something or not. Such binary variables are often thought to be indicators of one or more underlying latent variables like, for instance, ability or attitude. \citet{Bartholomew_1999} classify latent variable models into four different classes according to the respective natures of the manifest and latent variables (cf.\ Table~\ref{table:1.1}). Note that latent trait analysis is termed item response theory (IRT) in the field of educational testing and psychological measurement. 
\begin{table}[ht]
\centering
\caption{The classification of latent variable methods used by \cite{Bartholomew_1999}.}
\resizebox{0.9\columnwidth}{!}{ 
\begin{tabular*}{1\textwidth}{@{\extracolsep{\fill}}l| l| l l}
\hline
\hline
&&\multicolumn{2}{c}{Manifest Variables} \\ 
\cline{3-4}
&&Metrical&Categorical\\ [0.5ex] 
\hline
\multirow{2}{*}{Latent Variables}&Metrical&Factor analysis&Latent trait analysis\\
&Categorical&Latent profile analysis&Latent class analysis\\
  [1ex]
\hline
\end{tabular*}}
\label{table:1.1}
\end{table}

Model-based clustering is a principled statistical approach for clustering, where the data are clustered using some assumed mixture modelling structure. A finite mixture model is a convex combination of a finite number of simple component distributions. Historically, the Gaussian mixture model has dominated the model-based clustering literature \citep[e.g.,][]{wolfe63,Banfield_1993,Celeux_1995,Fraley_2002,McNicholas_2008,McLachlan_2010}. However, very recent model-based clustering work has focused on mixtures of non-elliptical distributions \citep[e.g.,][]{lin10, lee13,vrbik12,vrbik14,Franczak_2012,murray14a,murray14b}. Mixture model approaches to data where some or all of the variables are discrete have also been considered \citep[e.g.,][]{hunt99, Mclachlan_2000, mclachlan04}. 

Model-based approaches for categorical data have received relatively little attention, and recent work on mixtures of latent trait models is summarized in Table~\ref{table:1.2}. \citet{Browne_2012} introduce a mixture of latent variables models for the model-based clustering of data with mixed type, and a data set with all binary variables fits within their modelling framework as a special case. They use the deterministic annealing approach to estimate the likelihood described in \citet{Zhou_2010}. This approach focuses on increasing the chance of finding the global maximum; however, Gauss-Hermite quadrature is required to approximate the likelihood.
\begin{table}[ht]
\caption[Model-based clustering work on discrete data]{Model-based clustering work on discrete data}
\resizebox{0.7\columnwidth}{!}{ 
  \begin{minipage}{\textwidth}
\begin{center}
\begin{tabular}{l l l l l}
\hline
Author&Response Function&Data Type &Likelihood Estimation\\
 \hline
\citet{Browne_2012} &Logit  \footnote{Binary data fits in the model as a special case.} & Binary/Continuous &Deterministic annealing\footnote{They use the deterministic annealing approach described by \citet{Zhou_2010}.}\\
\citet{Cagnone_2012} &Logit&Binary &Gauss-Hermite quadrature\\
\citet{Murphy_2013}&Logit &Binary&Variational EM \\
\citet{Muthen_2006}&Probit &Binary/ Ordered Categorical &Numerical integration\\
\citet{Vermunt_2007}&Logit &Multilevel Binary/ Ordered Categorical&Numerical integration\\
[1ex]
\hline
\end{tabular}
\end{center}
\end{minipage}}
\label{table:1.2}
\end{table}

\citet{Murphy_2013} propose a mixture of latent trait analyzers (MLTA) for model-based clustering of binary data, wherein a categorical latent variable identifies groups of observations and a latent trait is used to accommodate within cluster dependency. They consider a lower bound approximation to the log-likelihood. This approach is easy to implement and converges quickly in comparison with other numerical approximations to the likelihood. However, mixture of latent trait models become highly parameterized when applied to high-dimensional binary data, particularly when the data come from several different groups and the continuous latent variable is high-dimensional. 

A mixture of item response models (\Citealp{Muthen_2006}; \Citealp{Vermunt_2007}) has very similar structure to the latent trait mixture models; however, it is highly parameterized, uses a probit structure, and numerical integration is required to compute the likelihood. Thus, it can be difficult to apply to large heterogeneous data sets in practice. A similar approach has also been discussed by \citet{Cagnone_2012}, who use Gauss-Hermite quadrature to approximate the likelihood. In addition, they assume a semi-parametric distributional form for the latent variables by adding extra parameters to the model. 

Multilevel mixture item response models \citep{Vermunt_2007} can be used to cluster repeatedly sampled binary data. These models focus on univariate traits because of the number of parameters in the model and the use of quadrature methods for the numerical integration of continuous latent variables. Accordingly, multilevel mixture item response models are not suitable for analyzing large data sets with underlying high-dimensional latent trait structure. 

For these reasons, we propose two different mixtures of latent traits models with common slope parameters for model-based clustering of binary data: a general model that supposes that the dependence among the response variables within each observation is wholly explained by a $d$ dimensional continuous latent variable in each group, and an exclusive model for repeatedly sampled data that supposes the response function in each group is composed of two continuous latent variables by adding a blocking latent variable. The proposed family of mixture of latent trait models with common slope parameters (MCLT) is a categorical analogue of a mixture of common factor analyzers model \citep{McLachlan_2010}. 
The MCLT model enables us to reduce the number of free parameters considerably in estimating the slope. Moreover, it facilitates a low-dimensional visual representation of the clusters with posterior means of the continuous latent variables corresponding to the observed data. The model with a blocking latent variable can potentially reduce known variability of repeatedly sampled data among groups; accordingly, we can be more accurate about group identification. 

In the mixture of latent traits model, the likelihood function involves an integral that is intractable. In this work, we propose using a variational approximation of the likelihood, as proposed by \citet{Jaakkola_2000}, \citet{Tipping_1999} and \citet{Attias_2000}, considered a latent variable density model. For a fixed set of values for the variational parameters, the transformed problem has a closed-form solution, providing a lower bound approximation to the log-likelihood. The variational parameters are optimized in a separate step.

The general model is demonstrated on  a U.S.\ Congressional Voting data set \citep{Bache_2013} and the model for clustered data is applied to a data set describing the sensory properties of orange juice \citep{Lee_2013}. We compare our approach to the MLTA approach proposed by \citet{Murphy_2013}. 

The remainder of this paper is organized as follows. In Section~\ref{sec:method}, we propose a mixture of latent trait analyzers model with common slope parameters. The data simulations are presented in Section~\ref{sec:Simulation}. Our approach is then applied to two real data sets (Section~\ref{sec:Application}), and we conclude with a summary and suggestions for future work (Section~\ref{sec:conclusion}). 

\section{Mixture of Latent Trait Models with Common Slope Parameters}\label{sec:method}
\subsection{Overview}\label{method_overview}
The MCLT approach restricts the MLTA model by assuming that all latent traits have a set of common slope parameters $\boldsymbol{W}=(\boldsymbol{w}_1,\ldots, \boldsymbol{w}_m)$, for $M$ binary response variables and a $d$-dimensional continuous latent variable $\boldsymbol{Y}$ comes from $g$ different components, where $\boldsymbol{Y}_{ng}\sim\text{MVN}(\boldsymbol{\mu}_g, \boldsymbol{\Sigma}_g)$. Thus, the MCLT model is a mixture model for binary data that reduces the number of parameters to a manageable size; still, each latent trait has a different effect in each group. It also facilitates low-dimensional visual representation of components with posterior means of the continuous latent variables corresponding to the observed data.

Similar to MLTA model, we assume that each observation $\boldsymbol x_n$ ($n=1,\ldots, N$) comes from one of the $G$ components and we use $\boldsymbol z_n=(z_{n1},\ldots, z_{nG})$ to identify the group membership, where $z_{ng}=1$ if observation $n$ is in component $G$ and $z_{ng}=0$ otherwise. We assume that the conditional distribution of $\boldsymbol x_n$ in group~$g$ is a latent trait model. Therefore, the MCLT model takes the form,
\begin {equation} \label{eq:3.1}
p(\boldsymbol x_n)=\sum_{g=1}^G\eta_gp(\boldsymbol x_n|z_{ng}=1)=\sum_{g=1}^G\eta_g\int \limits_{\boldsymbol{\mathcal Y}_{ng}} p(\boldsymbol{x}_n|\boldsymbol{y}_{ng}, z_{ng}=1)p(\boldsymbol{y}_{ng})d\boldsymbol y_{ng},
\end{equation}
where
\begin {equation*}
 p(\boldsymbol{x}_n|\boldsymbol{y}_{ng}, z_{ng}=1)=\prod_{m=1}^M[\pi_{mg}(\boldsymbol{y}_{ng})]^{x_{nm}}[1-\pi_{mg}(\boldsymbol{y}_{ng})]^{1-x_{nm}},
\end{equation*}
and the response function for each categorical variable in each group is
\begin {equation} \label{eq:3.3}
\pi_{mg}(\boldsymbol{y}_{ng})=p(x_{nm}=1|\boldsymbol{y}_{ng}, z_{ng}=1)=\frac{1}{1+\exp\{-\boldsymbol{w}_m'\boldsymbol{y}_{ng}\}},
\end{equation}
where $\boldsymbol w_m$ is the common model parameter and the latent variable $\boldsymbol Y_{ng}\sim \text{MVN}(\boldsymbol{\mu}_g, \boldsymbol\Sigma_g)$.

The complete-data log-likelihood is then given by
\begin {equation} \label{eq:3.4}
l=\sum_{n=1}^N\log\left[\sum_{g=1}^G \eta_g \int \limits_{\boldsymbol{\mathcal Y}_{ng}} \prod_{m=1}^M \, p(x_{nm}|\boldsymbol{y}_{ng}, z_{ng}=1)p(\boldsymbol{y}_{ng})d\boldsymbol{y}_{ng}\right].
\end{equation}
Therefore, the model is a finite mixture model in which the $g$th component latent variable $\boldsymbol{Y}_{ng}$ is $\text{MVN}(\boldsymbol{\mu}_g, \boldsymbol{\Sigma}_g)$ and the mixing proportions are $\eta_1, \eta_2, \ldots, \eta_G$.

\subsection{MCLT with Block Effect}\label{sec:blocking}
 A specific model with block effect can be used for the analysis of clustered data. Clustered data arise, for example, in research designs where a sample of clusters is repeatedly assessed or in educational settings where pupils are clustered within schools. The outcomes stemming from the same cluster tend to be more homogeneous than outcomes stemming from different clusters; accordingly, the outcomes within a cluster are likely to be correlated. These dependencies are taken into account via a response function with a blocking latent variable. 

Suppose that each observation $\boldsymbol{x}_{ij}$ is the j$th$ observed outcome of cluster $i$ $(i=1,\ldots, I; \; j=1,\ldots, J)$, and the cluster-specific parameters $s_{ij}$ are assumed to explain all dependencies that are due to inter-cluster variability. Thus, the response function for each group is given by
\begin{equation*}
\begin{split}
\pi_{mg}(\boldsymbol{y}_{ijg},s_{ij})&=p(x_{ijm}=1|\boldsymbol{y}_{ijg}, s_{ij}, z_{ijg}=1)
=\frac{1}{1+\exp\{-(\boldsymbol{w}_m'\boldsymbol{y}_{ijg}+\beta_ms_{ij})\}},
\end{split}
\end{equation*}
where $\boldsymbol{w}_m$ and $\beta_m$ are the model parameters. In addition, it is assumed that the blocking latent variable $S_{ij}\sim\text{N}(b_i, \sigma_i^2)$.
The model follows naturally:
\begin {equation*}
\begin{split}
p(\boldsymbol{x}_{ij})&=\sum_{g=1}^G\eta_gp(\boldsymbol{x}_{ij}|z_{ijg}=1)
=\sum_{g=1}^G\eta_g\int\limits _{\mathbb {R}}\int\limits_{\boldsymbol{\mathcal Y}_{ijg}}p(\boldsymbol{x}_{ij}|\boldsymbol{y}_{ijg}, s_{ij}, z_{ijg}=1)p(\boldsymbol{y}_{ijg})p(s_{ij})d\boldsymbol{y}_{ijg}ds_{ij},
\end{split}
\end{equation*}
and the log likelihood can be written
\begin{equation*} \label{eq:3.7}
l=\sum_{i=1}^I \sum_{j=1}^J \log\left[\sum_{g=1}^G \eta_g \int\limits_{\mathbb{R}} \int \limits_{\boldsymbol{\mathcal Y}_{ijg}}\prod_{m=1}^M \, p(x_{ijm}|\boldsymbol{y}_{ijg}, s_{ij}, z_{ijg}=1)p(\boldsymbol{y}_{ijg})p(s_{ij})d\boldsymbol{y}_{ijg}ds_{ij}\right].
\end{equation*}

The MCLT model with block effect is closely related to the multilevel mixture item response models \citep{Vermunt_2007, Ng_2006}. \citet{Vermunt_2007} assumes the latent variables at each level can be continuous, discrete, or both. The MCLT model with block effect is one of the special cases: the lower-level latent variables are combinations of discrete and continuous, with continuous random effects at the higher level. However, we focus on a multivariate trait parameter and the use of common slope parameter considerably reduces the number of free parameters in the model. To the best of our knowledge, this is the first work taking a close look at this particular case.

\subsection{Gaussian Parsimonious Mixture Models}\label{sec:GPCM}

Following \citet{Banfield_1993} and \citet{Celeux_1995}, we consider a parametrization of the covariance matrices $\boldsymbol{\Sigma}_1,\ldots,\boldsymbol{\Sigma}_g$ of the component densities. The parametrization of the component covariance matrices via eigenvalue decomposition is
$$\boldsymbol{\Sigma}_g=\lambda_g\boldsymbol{Q}_g\boldsymbol{A}_g\boldsymbol{Q}_g',$$
where $\lambda_g=|\boldsymbol{\Sigma}_g|^{\frac{1}{d}}$, $\boldsymbol{Q}_g$ is the matrix of eigenvectors of $\boldsymbol{\Sigma}_g$, and $\boldsymbol{A}_g$ is a diagonal matrix, such that $|\boldsymbol{A}_g|=1$, with the normalized eigenvalues of $\boldsymbol{\Sigma}_g$ on the diagonal in a decreasing order. 
The parameter $\lambda_g$ determines the volume of the $g$th cluster, $\boldsymbol{Q}_g$ its orientation, and $\boldsymbol{A}_g$ its shape. We write $\boldsymbol{\Sigma}_g=\lambda_g\boldsymbol{B}_g$, where $\boldsymbol{B}_g$ is a diagonal matrix with $|\boldsymbol{B}_g|=1$. This particular parametrization gives rise to four models (corresponding to component covariance structures $\lambda \boldsymbol{B}$, $\lambda_g\boldsymbol{B}$, $\lambda \boldsymbol{B}_g$, and $\lambda_g \boldsymbol{B}_g$, respectively). By assuming spherical shapes, namely $\boldsymbol{A}_g=\boldsymbol{I}$, another two parsimonious models are available: $\lambda \boldsymbol{I}$ and $\lambda_g\boldsymbol{I}$. Finally, the 14 parameterizations in Table~\ref{table:3.2} are considered; note that the corresponding Gaussian mixture models make up the GPCM family of \citet{Celeux_1995}.
\begin{table} [!ht] 
\caption{Fourteen parameterizations of $\boldsymbol{\Sigma_g}$ and the associated number of free parameters.}
\label{table:3.2}
\centering
\scalebox{0.8}{
 \begin{minipage}{\textwidth}
\begin{tabular}[b]{l l l r}
\hline
&$\boldsymbol{\Sigma}_g$&Vol/Shape/Orientation\footnote{``E'' represents ``equal" and ``V" represents ``variable".}&Number of free parameters \\ [0.1ex] 
\hline
1&$\lambda \boldsymbol{Q} \boldsymbol{A}\boldsymbol{Q}'$&EEE&$G-1+d\left(M+G\right)+d(d+1)/2-d^2$ \\
2&$\lambda_g \boldsymbol{Q} \boldsymbol{A} \boldsymbol{Q}'$&VEE&$G-1+d\left(M+G\right)+d(d+1)/2+G-1-d^2$ \\
3&$\lambda \boldsymbol{Q} \boldsymbol{A}_g \boldsymbol{Q}'$&EVE& $G-1+d\left(M+G\right)+d(d+1)/2+(G-1)(d-1)-d^2$ \\
4&$\lambda_g \boldsymbol{Q} \boldsymbol{A}_g \boldsymbol{Q}'$ &VVE&$G-1+d\left(M+G\right)+d(d+1)/2+(G-1)d-d^2$ \\
5& $\lambda \boldsymbol{Q}_g \boldsymbol{A} \boldsymbol{Q}'_g$&EEV&$G-1+d\left(M+G\right)+G( d(d+1)/2)-(G-1)d-d^2$\\
6&$\lambda_g \boldsymbol{Q}_g \boldsymbol{A} \boldsymbol{Q}'_g$&VEV&$G-1+d\left(M+G\right)+G(d(d+1)/2)-(G-1)(d-1)-d^2$\\
7&$\lambda \boldsymbol{Q}_g\boldsymbol{A}_g\boldsymbol{Q}'_g$&EVV&$G-1+d\left(M+G\right)+G(d (d+1)/2)-(G-1)-d^2$\\
8&$\lambda_g\boldsymbol{Q}_g\boldsymbol{A}_g\boldsymbol{Q}'_g$&VVV&$G-1+d\left(M+G\right)+G( d(d+1)/2)-d^2$\\
9&$\lambda \boldsymbol{B}$&EEI&$G-1+d\left(M+G\right)+d-d^2$\\
10&$\lambda_g\boldsymbol{B}$&VEI&$G-1+d\left(M+G\right)+G+d-1-d^2$\\
11&$\lambda \boldsymbol{B}_g$&EVI&$G-1+d\left(M+G\right)+Gd-G+1-d^2$\\
12&$\lambda_g \boldsymbol{B}_g$&VVI&$G-1+d\left(M+G\right)+G d-d^2$\\
13&$\lambda \boldsymbol{I}$&EII&$G-1+d\left(M+G\right)+1-d^2$\\
14&$\lambda_g \boldsymbol{I}$&VII&$G-1+d\left(M+G\right)+G-d^2$\\
\hline
\end{tabular}\end{minipage}}
\end{table}

In Table~\ref{table:3.3}, we list the number of parameters to be estimated for the MLTA \citep{Murphy_2013}, parsimonious MLTA (PMLTA), and MCLT approaches for $M=50,100$, $d=2$, and $G=2,5$. For example, when we cluster $M=100$ dimensional data into $g=2$ groups using a $d=2$ dimensional latent variable, the MLTA model requires 599 parameters to be estimated, while MCLT needs at most $207$ parameters. Moreover, as the number of components grows from $2$ to $5$, the number of parameters grows almost twice as large as before, even for the PMLTA model, but the number of parameters for MCLT remains almost the same. 
\begin{table} [h!] 
\caption{The number of free parameters in models for three mixture of latent trait models.}
\label{table:3.3}
\centering
\scalebox{0.8}{
 \begin{minipage}{\textwidth}
\centering
\begin{tabular}[b]{l l l l r}
\hline
Model&$G$&$M$&$d$&Number of free parameters \\ [0.1ex] 
\hline
\multirow{4}{*}{MLTA}&$2$&$50$&$2$&299 \\
&$5$&$50$&$2$& 749\\
&$2$&$100$&$2$& 599\\
 &$5$&$100$&$2$&1499 \\ [1ex]

\multirow{4}{*}{PMLTA}&$2$&$50$&$2$&200\\
&$5$&$50$&$2$& 353\\
&$2$&$100$&$2$&400 \\
 &$5$&$100$&$2$& 703\\ [1ex]

\multirow{4}{*}{MCLT\footnote{There are 14 different models for each combination of $G$, $d$ and $M$ (Table~\ref{table:3.2}).}}&$2$&$50$&$2$&102\footnote{The minimum number of free parameters is calculated by using the coviriance structure $\lambda \boldsymbol{I}$.}--107\footnote{The maximum number of free parameters is calculated by using the coviriance structure $\lambda_g\boldsymbol{Q}_g\boldsymbol{A}_g\boldsymbol{Q}'_g$.}\\
&$5$&$50$&$2$& 111--125\\
&$2$&$100$&$2$& 202--207\\
 &$5$&$100$&$2$& 211--225\\
 [1ex]
\hline
\end{tabular}
\end{minipage}}
\end{table}

\subsection{Interpretation of Model Parameters} \label{sec:model_interpretation}
The interpretation of the model parameter can be exactly as in MLTA and IRT models. In the finite mixture model, $\eta_g$ is the probability of an observation sampling from the group $g$. The characteristics of component~$g$ are determined by a common slope $\boldsymbol{w}_m$, and by the hyperparameters of the latent variable $\boldsymbol{Y}_{ng}$. In the geometric interpretation of the multivariate normal distribution, the equidensity contours of a non-singular multivariate normal distribution are ellipsoids centred at the mean $\boldsymbol{\mu}$. The directions of the principal axes of the ellipsoids are given by the eigenvectors of the covariance matrix $\boldsymbol{\Sigma}$. If $\boldsymbol{Y} \sim\text{N}(\boldsymbol{\mu}, \boldsymbol{\Sigma})$, then we have $\boldsymbol{Y} \sim \boldsymbol{\mu}+\boldsymbol{\Sigma}^\frac{1}{2} \text{N}(\boldsymbol{0}, \boldsymbol{I})$. Thus, the response function in Equation~\ref{eq:3.3} can be written
\begin{equation}\label{eq:3.8}
\pi_{mg}(\boldsymbol{\tau}_{n})=p(x_{nm}=1|\boldsymbol{y}_{ng}, z_{ng}=1)=\frac{1}{1+\exp\{-(\boldsymbol{w}_m'\boldsymbol{\mu}_g+\boldsymbol{w}_m'\boldsymbol{\Sigma}_g^\frac{1}{2} \boldsymbol{\tau}_{n})\}},
\end{equation}
where $\boldsymbol{Y}_{ng}\sim\text{N}(\boldsymbol{\mu}_g, \boldsymbol{\Sigma}_g)$ and $\boldsymbol{\tau}_n \sim\text{N}(\boldsymbol{0},\boldsymbol{I})$.

Because we have $\boldsymbol{\tau}_n \sim\text{N}(\boldsymbol{0},\boldsymbol{I})$, the value $\pi_{mg}(0)$ can be used to examine the probability that the median individual in group $g$ has a positive response for the variable~$m$, 
\begin{equation}\label{eq:3.9}
\pi_{mg}(0)=p(x_{nm}=1|\boldsymbol{\tau}_n=0, z_{ng}=1)= \frac{1}{1+\exp\{-\boldsymbol{w}_m'\boldsymbol{\mu}_g\}}.
\end{equation}
%
Moreover, the mean $\boldsymbol{\mu}_g$ and covariance matrix $\boldsymbol{\Sigma}_g$ of component~$g$ can be used to provide 
low-dimensional plots of the cluster. 

In the MCLT model with block effect, we can write $\pi^*_{mg}(0)$, which is the $\pi_{mg}(0)$ in \eqref{eq:3.9} adjusted by block effect, as
\begin{equation*}
\pi^*_{mg}(0)=\frac{1}{I}\left[\sum_{i=1}^I \frac{1}{1+\exp(-(\boldsymbol{w}{'}_m\boldsymbol{\mu}_g+\beta{'}_m s_i))}\right],
\end{equation*}
where $I$ is the number of blocks.

\subsection{Related Models} \label{sec:related}
There are other statistical models that share a lot of common characteristics with our model. 
The MCLT model can be treated as a categorical version of a mixture of common factor analyzers (MCFA) model \citep{McLachlan_2010}. The MCFA model employs constraints on the $g$ component means and covariance matrices, i.e., 
\begin{equation*}
\boldsymbol{\mu}_g=\boldsymbol{A}\boldsymbol{\upsilon}_g, \qquad \text{and} \qquad
\boldsymbol{\Sigma}_g=\boldsymbol{A}\boldsymbol{\psi}_g\boldsymbol{A}'+\boldsymbol{D}.
\end{equation*}

A common factor loading matrix $\boldsymbol{A}$ is analogous to a common trait parameter in MCLT. The component covariance matrix is analogous to the covariance matrix of the response function's posterior distribution in each group. The component mean is identical to the mean of the response function's posterior distribution. Of course, the mixing proportions take a same role in both models. 

\citet{Davier_2007} consider a mixture Rasch model \citep{Rasch_1960} that is equivalent to the parsimonious model of \citet{Murphy_2013}. The model is given by
\begin{equation*}
P(x_{nm}=1| \boldsymbol{q}_m, \boldsymbol{\beta}_n, \sigma_m, z_{ng}=1)=\frac{\exp(\boldsymbol{q}_m'\boldsymbol{\beta}_n-\sigma_{mg})}{1+\exp(\boldsymbol{q}_m'\boldsymbol{\beta}_n-\sigma_{mg})},
\end{equation*}
where $\boldsymbol{q}_m$ are variable-specific parameters, $\boldsymbol{\beta}_n$ is the $d$-dimensional ability parameter, and $\sigma_{mg}$ is the difficulty parameter. 

Many other versions of mixture models with response functions have been proposed for analyzing binary data, including the mixed latent trait model \citep{Uebersax_1999}, the latent class factor analysis (LCFA) model \citep{Vermunt_2005}, and a range of mixture item response models \citep{Bolt_2001, Muthen_2006, Vermunt_2007, Vermunt_2008}. A key difference between our model and other mixture models is that we focus on a mixture of multivariate latent variables, which allows us to provide low-dimensional plots of the clusters. In addition, we implement a variational approximation for parameter estimation of latent trait models that provides a computationally efficient means of model fitting.

\subsection{Variational Approximation} \label{sec:variational approx}
\citet{Jaakkola_2000} introduced a variational approximation for the predictive likelihood in a Bayesian logistic regression model and also briefly considered the ``dual'' problem, which is closely related to the latent trait model. It obtains a closed form approximation to the posterior distribution of the parameters within a Bayesian framework. Their method is based on a second order Taylor series expansion of the logistic function around a point
\begin{equation*}  \label{eq:2.9}
p(\boldsymbol{x}_{nm}=1|\boldsymbol{y}_{ng}, \, z_{ng}=1)=\frac{\exp\{\boldsymbol w'_m\boldsymbol{y}_{ng}\}}{1+\exp\{\boldsymbol w'_m\boldsymbol{y}_n\}}=(1+\exp\{-\boldsymbol w'_m\boldsymbol{y}_n\})^{-1}.
\end{equation*}
where $\xi_{nmg}\neq 0$ for all $m=1,...,M$. Now, the lower bound of each term in the log-likelihood is given by, 
\begin{equation}  \label{eq:2.10}
L(\boldsymbol{\xi}_{ng})=\log(\tilde{p}(\boldsymbol{x}_n|\boldsymbol{\xi}_{ng})=\log\left(\int \prod_{m=1}^M \tilde{p}(x_{nm}|\boldsymbol{y}_{ng}, z_{ng}=1, \xi_{nmg}) p(\boldsymbol{y}_{ng})\, d\boldsymbol{y}_{ng}\right),
\end{equation}
where
\begin{equation*}\begin{split}  \label{eq:2.11}
&\tilde{p}(x_{nm}|\boldsymbol{y}_{ng}, z_{ng}=1,\xi_{nmg})=\sigma(\xi_{nmg})\exp\left\{\frac{A_{nmg}-\xi_{nmg}}{2}+ \lambda(\xi_{nmg})(A_{nmg}^2-\xi_{nmg}^2)\right\},\\
&A_{nmg}=(2x_{nm}-1)(\boldsymbol{w}'_{m}\boldsymbol{y}_{ng}),\
\lambda(\xi_{nmg})=\frac{1}{2\xi_{nmg}}\left[\frac{1}{2}-\sigma(\xi_{nmg})\right],\
\sigma(\xi_{nmg})=(1+\exp\{-\xi_{nmg}\})^{-1}.
\end{split}\end{equation*}
This approximation is used to obtain a lower bound for the log-likelihood. A variational EM algorithm \citep{Tipping_1999} can then be used to obtain parameter estimates that maximize this lower bound. 

\subsection{Model Fitting}\label{sec:fitting}
When fitting the MCLT model, the integral in the log-likelihood (\ref{eq:3.4}) is intractable. Here we illustrate how to use a variational EM algorithm to obtain the approximation of the likelihood: 
\begin{enumerate}
  \item E-Step: estimate $z_{ng}^{(t+1)}$ using
\begin{equation*}
z_{ng}^{(t+1)}=\frac{\eta_g^{(t)} \exp\{L(\boldsymbol{\xi}_{ng}^{(t)})\}}{\sum_{g=1}^G \eta{'}_{g}^{(t)} \exp\{L(\boldsymbol{\xi}{'}_{ng}^{(t)})\}}.
 \end{equation*}
\item M-Step: estimate $\eta_g^{(t+1)}$ using
\begin{equation*}
\eta_g^{(t+1)}=\frac{1}{N}\sum_{n=1}^Nz_{ng}^{(t+1)}.
\end{equation*}

\item Estimate the lower bound of log-likelihood via variational parameter $\xi_{nmg}$:
\begin{enumerate}
\item E-Step: we approximate the latent posterior statistics for $p(\boldsymbol{y}_{ng}|\boldsymbol{x}_n, z_{ng}^{(t+1)}=1)$ by its variational lower bound $\underline{p}(\boldsymbol{y}_{ng}|\boldsymbol{x}_n, z_{ng}^{(t+1)}=1, \boldsymbol{\xi}_{ng}^{(t)} )$, which is a $\text{N}(\boldsymbol{\upsilon}_{ng}^{(t+1)},\boldsymbol{\varphi}_{ng}^{(t+1)})$ density, where
\begin{equation*}
\begin{split}
&(\boldsymbol{\varphi}^{-1}_{ng})^{(t+1)}=(\boldsymbol{\Sigma}_{g}^{-1})^{(t)}-2\sum_{m=1}^M\lambda(\xi_{nmg}^{(t)})\, \boldsymbol{w}_{m}^{(t)}\boldsymbol{w}{'}_{m}^{(t)},\\
&\boldsymbol{\upsilon}_{ng}^{(t+1)}=\boldsymbol{\varphi}_{ng}^{(t+1)}\left[(\boldsymbol{\Sigma}_{g}^{-1})^{(t)}\boldsymbol{\mu}_{g}^{(t)} +\sum_{m=1}^M\left(x_{nm}-\frac{1}{2}\right)\boldsymbol{w}_{m}^{(t)}\right],
\end{split}
\end{equation*}
where $\sigma(\xi_{nmg})=(1+\exp\{-\xi_{nmg}\})^{-1}$, $\lambda(\xi_{nmg})=(\frac{1}{2}-\sigma(\xi_{nmg}))/2\xi_{nmg}$.

\item M-Step: optimize the variational parameter $\xi_{nmg}^{(t+1)}$. Owing to the EM formulation, each update for $\xi_{nmg}$ corresponds to a monotone improvement to the posterior approximation. The update is
\begin{equation*}
(\xi_{nmg}^2)^{(t+1)}=\boldsymbol{w}{'}_{m}^{(t)}\mathrm{E}[\boldsymbol{y}_{ng}\boldsymbol{y}{'}_{ng}]\boldsymbol{w}_{m}^{(t)} ,
\end{equation*}
where the expectation is taken with respect to $\underline{p}(\boldsymbol{y}_{ng}|\boldsymbol{x}_n, z_{ng}^{(t+1)}=1, \boldsymbol{\xi}_{ng}^{(t)} )$, the variational posterior distribution based on the previous value of $\xi_{nmg}$. Thus, we have $\mathrm{E}[\boldsymbol{y}_{ng}\boldsymbol{y}{'}_{ng}]=\boldsymbol{\varphi}_{ng}^{(t+1)}+\boldsymbol{\upsilon}_{ng}^{(t+1)}\boldsymbol{\upsilon}{'}_{ng}^{(t+1)}$.
\item Update parameters $\boldsymbol{w}_m$, $\boldsymbol{\mu}_g$, and $\boldsymbol{\Sigma}_g$ based on the posterior distributions corresponding to the observations in the data set:\begin{equation*}
\begin{split}
&\boldsymbol{\Sigma}_{g}^{(t+1)}=\frac{1}{n_g}\sum_{n=1}^N z_{ng}^{(t+1)}\boldsymbol{\varphi}_{ng}^{(t+1)},\\
&\boldsymbol{\mu}_{g}^{(t+1)}=\frac{1}{n_g}\sum_{n=1}^N z_{ng}^{(t+1)}\boldsymbol{\upsilon}_{ng}^{(t+1)},
\end{split}
\end{equation*}
where $n_g=z_{1g}+\cdots + z_{Ng}$ and
\begin{equation*}
\begin{split}
\boldsymbol{w}_{m}^{(t+1)}=&-\left[2\sum_{g=1}^G \sum_{n=1}^N z_{ng}^{(t+1)}\,\lambda(\xi_{nmg}^{(t+1)})\,\,(\boldsymbol{\varphi}_{ng}^{(t+1)}+\boldsymbol{\upsilon}_{ng}^{(t+1)} \boldsymbol{\upsilon}{'}_{ng}^{(t+1)})\right]^{-1}\\
&\qquad\qquad\qquad\qquad\qquad\qquad\qquad\qquad\times \left[\sum_{g=1}^G \sum_{n=1}^N z_{ng}^{(i+1)} (x_{nm}-\frac{1}{2})\boldsymbol{\upsilon}_{ng}^{(i+1)}\right].
\end{split}
\end{equation*}
\item Obtain the lower bound of the log-likelihood at the expansion point $\xi_{ng}$:
\begin{equation*}
\begin{split}
L(\boldsymbol{\xi}_{ng}^{(t+1)})=&\sum_{m=1}^M\left[\log\sigma(\xi_{nmg}^{(t+1)})-\frac{\xi_{nmg}^{(t+1)}}{2}-\lambda(\xi_{nmg}^{(t+1)})(\xi^2_{nmg})^{(t+1)}\right]\\
&-\frac{\boldsymbol{\mu}{'}_{g}^{(t+1)}(\boldsymbol{\Sigma}_{g}^{-1})^{(t+1)}\boldsymbol{\mu}_{g}^{(t+1)}}{2}  +\frac{1}{2}\log\frac{|\boldsymbol{\varphi}_{ng}^{(t+1)}|}{|\boldsymbol{\Sigma}_{g}^{(t+1)}|}+\frac{\boldsymbol{\upsilon}{'}_{ng}^{(t+1)}(\boldsymbol{\varphi}_{ng}^{-1})^{(t+1)}\boldsymbol{\upsilon}_{ng}^{(t+1)}}{2},
\end{split}
\end{equation*}
and the log-likelihood:
\begin{equation*}
l^{(t)}\approx\sum_{n=1}^N\log\left[\sum_{g=1}^G\eta_g^{(t+1)}\exp\{L(\boldsymbol{\xi}_{ng}^{(t+1)})\}\right].
\end{equation*}
\end{enumerate}
\item Return to Step 1. 
\begin{enumerate}
\item The stopping criterion adopted here is a measure of lack of progress when all parameter estimates become stable and no further improvements can be made to the likelihood value.
$$|\theta^{(t+1)}-\theta^{(t)}|<0.01.$$
\item In our application to the voting data (Section~\ref{sec:Voting}), to facilitate comparison with the MLTA models, convergence of our variational EM algorithm is determined as in \cite{Murphy_2013}. They using a criterion based on the Aitken acceleration \citep{Aitken_1926}, stopping the algorithm when 
$$|l_A^{(t+1)}-l_A^{(t)}|\le 0.01,$$
where $$l_A^{(t+1)}=l^{(t)}+\frac{1}{1-a^{(t)}}(l^{(t+1)}-l^{(t)})$$ and $$a^{(t)}=\frac{l^{(t+1)}-l^{(t)}}{l^{(t)}-l^{(t-1)}}.$$ See \cite{bohning94} for details.
\end{enumerate}

\item Approximate the log-likelihood by using the Gauss-Hermite quadrature \citep{Bock_1981}.
\end{enumerate}

\subsubsection{Model Fitting with Block Effect}\label{sec:block_fit}

To fit model with block effect outlined in Section~\ref{sec:blocking}, we will need to re-derive the necessary expressions:
\begin{enumerate}
  \item E-Step: estimate $z_{ijg}^{(t+1)}$ with
\begin{equation*}
z_{ijg}^{(t+1)}=\frac{\eta_g^{(t)} \exp\{L(\boldsymbol{\xi}_{ijg}^{(t)})\}}{\sum_{g=1}^G \eta{'}_{g}^{(t)} \exp\{L(\boldsymbol{\xi}{'}_{ijg}^{(t)})\}}.
\end{equation*} 
\item M-Step: estimate $\eta_g^{(t+1)}$ using
\begin{equation*}
\eta_g^{(t+1)}=\frac{\sum_{i=1}^I \sum_{j=1}^Jz_{ijg}^{(t+1)}}{N}.
\end{equation*}

\item Estimate the likelihood
\begin{enumerate}
\item E-Step: estimate the latent posterior statistics for $\underline{p}(\boldsymbol{y}_{ijg}, s_{ijg}|\boldsymbol{x}_{ij}, z_{ijg}^{(t+1)}=1, \boldsymbol{\xi}_{ijg}^{(t)} )$ which is a $\text{N}(\hat{\boldsymbol{\upsilon}}_{ijg}^{(t+1)},\hat{\boldsymbol{\varphi}}_{ijg}^{(t+1)})$ density:
\begin{equation*}
\begin{split}
&{(\hat{\boldsymbol{\varphi}}_{ijg}^{-1})}^{(t+1)}=(\hat{\boldsymbol{\Sigma}}_{ijg}^{-1})^{(t)}-2\sum_{m=1}^M\lambda(\xi_{ijmg}^{(t)})\, \hat{\boldsymbol{w}}_{m}^{(t)}\hat{\boldsymbol{w}}{'}_{m}^{(t)},\\
&\hat{\boldsymbol{\upsilon}}_{ijg}^{(t+1)}=\hat{\boldsymbol{\varphi}}_{ijg}^{(t+1)}\left[(\hat{\boldsymbol{\Sigma}}_{ijg}^{-1})^{(t)}\hat{\boldsymbol{\mu}}_{ijg}^{(t)} +\sum_{m=1}^M(x_{ijm}-\frac{1}{2})\hat{\boldsymbol{w}}_{m}^{(t)}\right],
\end{split}
\end{equation*}

where $p(\boldsymbol{y}_{ijg}, s_{ijg})$ is a joint distribution of $p(\boldsymbol{y}_{ijg})$ and $p(s_{ijg})$, which is $\text{N}(\hat{\boldsymbol{\mu}}_{ijg}^{(t)}, \hat{\boldsymbol{\Sigma}}_{ijg}^{(t)})$ with $\hat{\boldsymbol{\mu}}_{ijg}^{(t)}= (\boldsymbol{\mu}_g^{(t)},b_i^{(t)})'$ and $$\hat{\boldsymbol{\Sigma}}_{ijg}^{(t)}= \left( \begin{array}{cc}
\boldsymbol{\Sigma}_g^{(t)}&0\\
0&(\sigma_i^{2})^{(t)}\\
\end{array}\right).$$

\item M-Step: optimize the variational parameter $\xi_{ijmg}^{(t+1)}$: 
$$(\xi_{ijmg}^{2})^{(t+1)}=\hat{\boldsymbol{w}}{'}_{m}^{(t)}\left[\hat{\boldsymbol{\varphi}}_{ijg}^{(t+1)}+\hat{\boldsymbol{\upsilon}}_{ijg}^{(t+1)}\hat{\boldsymbol{\upsilon}}{'}_{ijg}^{(t+1)}\right]\hat{\boldsymbol{w}}_{m}^{(t)}.$$

\item Update the parameters $\boldsymbol{w}_m$, $\beta_m$, $\boldsymbol{\mu}_g$, $\boldsymbol{\Sigma}_g$, $b_i$, and $\sigma_i^2$:
\begin{equation*}
\begin{split}
&\boldsymbol{\Sigma}_{g}^{(t+1)}=\frac{1}{n_g}\sum_{i=1}^I \sum_{j=1}^J z_{ijg}^{(t+1)}\hat{\boldsymbol{\varphi}}_{ijg}^{(t+1)},\qquad
\boldsymbol{\mu}_{g}^{(t+1)}=\frac{1}{n_g}\sum_{i=1}^I\sum_{j=1}^J z_{ijg}^{(t+1)}\hat{\boldsymbol{\upsilon}}_{ijg}^{(t+1)},\\
&(\sigma_{i}^{2})^{(t+1)}=\frac{1}{n_i}\sum_{g=1}^G \sum_{j=1}^J z_{ijg}^{(t+1)}\hat{\boldsymbol{\varphi}}_{ijg}^{(t+1)},\qquad
b_{i}^{(t+1)}=\frac{1}{n_i}\sum_{g=1}^G\sum_{j=1}^J z_{ijg}^{(t+1)}\hat{\boldsymbol{\upsilon}}_{ijg}^{(t+1)},
\end{split}
\end{equation*}
where $n_g=\sum_{i=1}^I \sum_{j=1}^Jz_{ijg}$, $n_i=\sum_{g=1}^G \sum_{j=1}^Jz_{ijg}$, and
\begin{equation*}
\begin{split}
\hat{\boldsymbol{w}}_{m}^{(t+1)}=&-\left(2\sum_{g=1}^G \sum_{i=1}^I\sum_{j=1}^J z_{ijg}^{(t+1)}\,\lambda(\xi_{ijmg}^{(t+1)})\,\,(\hat{\boldsymbol{\varphi}}_{ijg}^{(t+1)}+\hat{\boldsymbol{\upsilon}}_{ijg}^{(t+1)} \hat{\boldsymbol{\upsilon}}{'}_{ijg}^{(t+1)})\right)^{-1}\\
&\qquad\qquad\qquad\qquad\qquad\qquad\qquad\times \left(\sum_{g=1}^G \sum_{i=1}^I \sum_{j=1}^J z_{ijg}^{(t+1)} (x_{ijm}-\frac{1}{2})\hat{\boldsymbol{\upsilon}}_{ijg}^{(t+1)}\right),
\end{split}
\end{equation*}
where $\hat{\boldsymbol{w}}_m^{(t+1)}=(w_{m1}^{(t+1)},\ldots, w_{md}^{(t+1)},\beta_m^{(t+1)})'.$

\item Obtain the lower bound of the log-likelihood at the expansion point $\xi_{ijg}$:
\begin{equation*}
\begin{split}
L(\boldsymbol{\xi}_{ijg}^{(t+1)})=&\sum_{m=1}^M\left[\log\sigma(\xi_{ijmg}^{(t+1)})-\frac{\xi_{ijmg}^{(t+1)}}{2}-\lambda(\xi_{ijmg}^{(t+1)})(\xi_{ijmg}^2)^{(t+1)}\right]\\
&-\frac{\hat{\boldsymbol{\mu}}{'}_{ijg}^{(t+1)}(\hat{\boldsymbol{\Sigma}}_{ijg}^{-1})^{(t+1)}\hat{\boldsymbol{\mu}}_{ijg}^{(t+1)}}{2}  +\frac{1}{2}\log\frac{|\hat{\boldsymbol{\varphi}}_{ijg}^{(t+1)}|}{|\hat{\boldsymbol{\Sigma}}_{ijg}^{(t+1)}|}+\frac{\hat{\boldsymbol{\upsilon}}{'}_{ijg}^{(t+1)}(\hat{\boldsymbol{\varphi}}_{ijg}^{-1})^{(t+1)}\hat{\boldsymbol{\upsilon}}_{ijg}^{(t+1)}}{2},
\end{split}
\end{equation*}
and the log-likelihood:
\begin{equation*}
l^{(t)}\approx\sum_{i=1}^I\sum_{j=1}^J \log\left[\sum_{g=1}^G\eta_g^{(t+1)}\exp\{L(\boldsymbol{\xi}_{ijg}^{(t+1)})\}\right].
\end{equation*}
\end{enumerate}
\end{enumerate}

\subsection{Model Selection}\label{sec:selection}

We use the Bayesian information criterion \citep[BIC;][]{Schwarz_1978} as a criterion for model selection, 
\begin{equation}
\text{BIC}=-2l+k\log N,
\end{equation}
where $l$ is the maximized log likelihood, $k$ is the number of free parameters to be estimated in the model, and $N$ is the number of the observations. Within the framework of MCLT models, the values of number of components $G$, the dimension of the latent variable $Y$ (i.e., $d$), and the structure of the connivance matrices $\boldsymbol{\Sigma}_g$ need to be determined. Models with lower values of BIC are preferable. The BIC value could be overestimated using the variational approximation of log-likelihood, which is always less than or equal to the true value. For model selection purposes, we calculate maximized log-likelihood using Gauss-Hermite quadrature after convergence is attained.

For high-dimensional binary data, particularly when the number of observations $n$ is not very large relative to their dimension $m$, it is common to have a large number of patterns with small observed frequency. We cannot use a $\chi^2$ test to check the goodness of the model fit. The analysis of the groups in the selected model can be used to interpret the model. The adjusted Rand index \citep[ARI;][]{Rand_1971,Hubert_1985} can be used to assess the model performance. The ARI is the corrected-for-chance version of the Rand index. The general form is $$\frac{\text{index}-\text{expected index}}{\text{maximum index}-\text{expected index}},$$ which is bounded above by 1, and has expected value 0. Intuitively, an ARI value of~1 corresponds to perfect agreement, and a value of 0 would be expected under random classification.

\subsection{Model Identifiability }\label{sec:identifiability}
The identifiability of our model depends on the identifiability of the latent trait part as well as the identifiability of the mixture models. The identifiability of mixture models has been discussed in \citet{Mclachlan_2000}. \citet{Bartholomew_1999} give a detailed explanation of model identifiability in the latent trait models context. 

The slope parameters $\boldsymbol W_g$ are only identifiable with $d\times d$ constraints. This is important when determining the number of free parameters in the model (Table~\ref{table:3.2}). 

In addition, \citet{Murphy_2013} mention that model identifiability holds if the observed information matrix is full rank. This can be checked using empirical methods as possible non-identifiability can be identified through high standard errors of the parameter estimates and inconsistency between the maximized likelihood values from different random starts. 

These checks for identifiability are carried out in our simulation studies (Section~\ref{sec:Simulation}) and empirical examples (Section~\ref{sec:Application}). 

\subsection{Computational Aspects}\label{sec:compute}
We initialize the categorical latent variables $\boldsymbol{z}_n$ $(n=1,\ldots, N)$ by randomly assigning each observation to one of the $G$ groups. The variational parameters $\xi_{nmg}$ $(n=1, \ldots, N, \,\,\, m=1,\ldots, M, \,\,\, g=1,\ldots,G)$ are initialized to equal 20, which leads the initial approximation to the conditional distribution to 0. The model parameters are initialized by generating random numbers from a $\text{N}(0,1)$ distribution. The prior means of the latent variable $\boldsymbol{Y}_{ng}$ $(n=1, \ldots, N \,\,\, g=1,\ldots,G)$ are initialized by random generated number from a $N(0,1)$. We use $d$-dimensional identity matrices as the initial prior covariance matrices of $\boldsymbol{Y}_{ng}$ $(n=1, \ldots, N \,\,\, g=1,\ldots,G)$. In addition, the prior mean $\boldsymbol{b}=(b_1,\ldots, b_i)$ and the prior variance $\boldsymbol{\sigma}^2=(\sigma^2_1,\ldots,\sigma^2_i)$ of the blocking latent variable are set by generating random number from a $N(0,1)$. We start with ten random initializations of the algorithm and select the model with the lowest BIC.

The use of the variational EM algorithm leads us to an exactly solvable EM algorithm of a latent variable density model that guarantees monotone improvement in the approximation to the likelihood. We also find that this procedure converges rapidly, i.e., only a few iterations are needed. 

\section{Simulation Studies}\label{sec:Simulation}
To illustrate the accuracy of the proposed MCLT model, we performed a simulation experiment on a 20-dimensional binary data set (i.e.,~$M=20$). Thus a comparison of approaches (MLTA vs.\ MCLT) can be carried out. The observations are generated from a MCLT model of the form given in Equation~\ref{eq:3.1} with a two-component mixture ($G=2, \eta_1=0.5$). The latent variables are two-dimensional multivariate normal distributions. The first component has mean $\boldsymbol \mu_1=(0,1)'$, while the second component has mean $\boldsymbol \mu_2=(3,3)'$. The covariance matrices take the form, $\boldsymbol \Sigma_g=\lambda \boldsymbol B_g$. We choose sample sizes $n\in \{100,250,500\}$, and run 100 simulations for each sample. 

Tables~\ref{table:4.1} and~\ref{table:4.2} present the value of true model parameters as well as their mean squared errors (MSE) for $n=100, 250, 500$.  The MSEs decrease with increasing sample size $n$.
\begin{table}[ht]
\centering
\caption[Standard Error of $\boldsymbol w_m$]{True values and the MSEs of $\boldsymbol w_m$, tabulated against $n$.}
\scalebox{0.7}{
\begin{tabular}{c c c c c c | c c c c c c}
\hline
Variable&Parameters&True&$n=100$&$n=250$&$n=500$&Variable&Parameters&True&$n=100$&$n=250$&$n=500$\\ [0.5ex] 
\hline
\multirow{2}{*}{M1}&$w_{11}$&-1.0&$0.09$&$0.07$&$0.05$&\multirow{2}{*}{M11}&$w_{111}$&0.9&$0.16$&$0.04$&$0.03$\\
&$w_{12}$&-0.7&$0.13$&$0.06$&$0.07$&&$w_{112}$&0.6&$0.05$&$0.04$&$0.04$\\  [1ex]

\multirow{2}{*}{M2}&$w_{21}$&-0.3&$0.20$&$0.13$&$0.08$&\multirow{2}{*}{M12}&$w_{121}$&-0.4&$0.05$&$0.05$&$0.03$\\

&$w_{22}$&1.0&$0.65$&$0.16$&$0.06$&&$w_{122}$&1.7&$0.31$&$0.15$&$0.05$\\ [1ex]

\multirow{2}{*}{M3}&$w_{31}$&0.88&$0.36$&$0.09$&$0.04$&\multirow{2}{*}{M13}&$w_{131}$&0.9&$0.34$&$0.06$&$0.01$\\

&$w_{32}$&0&$0.37$&$0.09$&$0.08$&&$w_{132}$&0.8&$0.04$&$0.03$&$0.00$\\ [1ex]

\multirow{2}{*}{M4}&$w_{41}$&-0.7&$0.01$&$0.00$&$0.00$&\multirow{2}{*}{M14}&$w_{141}$&1.5&$0.09$&$0.01$&$0.00$\\

&$w_{42}$&0.4&$0.06$&$0.04$&$0.04$&&$w_{142}$&0&$0.09$&$0.09$&$0.04$\\ [1ex]
\multirow{2}{*}{M5}&$w_{51}$&0.6&$0.04$&$0.02$&$0.01$&\multirow{2}{*}{M15}&$w_{151}$&1.6&$0.2$&$0.1$&$0.01$\\

&$w_{52}$&-0.4&$0.06$&$0.06$&$0.05$&&$w_{152}$&0.5&$0.06$&$0.03$&$0.03$\\ [1ex]
\multirow{2}{*}{M6}&$w_{61}$&-0.4&$0.02$&$0.01$&$0.01$&\multirow{2}{*}{M16}&$w_{161}$&-0.5&$0.3$&$0.1$&$0.03$\\
&$w_{62}$&0&$0.22$&$0.05$&$0.01$&&$w_{162}$&-0.7&$0.10$&$0.02$&$0.00$\\ [1ex]
\multirow{2}{*}{M7}&$w_{71}$&2&$0.04$&$0.02$&$0.00$&\multirow{2}{*}{M17}&$w_{171}$&-0.5&$0.02$&$0.02$&$0.00$\\
&$w_{72}$&0.4&$0.16$&$0.26$&$0.01$&&$w_{172}$&-0.7&$0.01$&$0.01$&$0.00$\\ [1ex]
\multirow{2}{*}{M8}&$w_{81}$&-0.5&$0.00$&$0.00$&$0.00$&\multirow{2}{*}{M18}&$w_{181}$&-1.0&$0.01$&$0.00$&$0.00$\\
&$w_{82}$&-0.4&$0.02$&$0.02$&$0.00$&&$w_{182}$&0.6&$0.01$&$0.01$&$0.00$\\ [1ex]

\multirow{2}{*}{M9}&$w_{91}$&-1&$0.01$&$0.00$&$0.00$&\multirow{2}{*}{M19}&$w_{191}$&0.0&$0.00$&$0.00$&$0.00$\\
&$w_{92}$&-0.7&$0.02$&$0.01$&$0.00$&&$w_{192}$&2.8&$0.01$&$0.03$&$0.00$\\ [1ex]

\multirow{2}{*}{M10}&$w_{101}$&0.7&$0.00$&$0.00$&$0.00$&\multirow{2}{*}{M20}&$w_{201}$&-1.5&$0.02$&$0.00$&$0.00$\\
&$w_{102}$&0.5&$0.00$&$0.00$&$0.00$&&$w_{202}$&-0.9&$0.02$&$0.01$&$0.00$\\ [1ex]
\hline
\end{tabular}}
\label{table:4.1}
\end{table} 
\begin{table}[ht]
\centering
\caption[Standard Error of $\boldsymbol \mu_g$]{True values and the MSEs of $\boldsymbol \mu_g$, tabulated against $n$.}
\begin{tabular}{c c c c c c }
\hline
&Parameters&True&$n=100$&$n=250$&$n=500$\\ [0.5ex] 
\hline
\multirow{2}{*}{Group 1}&$\mu_{11}$&0&$0.07$&$0.07$&$0.01$\\
&$\mu_{12}$&1&$0.17$&$0.06$&$0.02$\\  [1ex]
\multirow{2}{*}{Group 2}&$\mu_{21}$&3&$0.09$&$0.10$&$0.03$\\
&$\mu_{22}$&3&$0.04$&$0.01$&$0.01$\\ [1ex]

\hline
\end{tabular}
\label{table:4.2}
\end{table} 

In Table~\ref{table:4.3}, we present a comparison of two different approaches on ARI from the clustering results for $n=100, 250, 500$. Each couplet in Table~\ref{table:4.3} shows the average ARI and its standard error of ARIs from 100 simulations. With the MCLT approach, the average ARI is $0.64$ with a standard error 0.008 for sample size as small as 100 on a 20-dimensional binary data; and a stable clustering result occurs when sample size reaches $250$. On the other hand, the average ARI is $0.48$ with a standard error of $0.03$ for the MLTA approach when $n=100$; and a stable clustering result only occurs when sample size is $500$. The average ARIs using MCLT approach are at least as good as those using MLTA approach for all sample sizes. 
\begin{table}[ht]
\centering
\caption[A comparison of MLTA and MCLT]{A comparison of two different approaches on ARI and their standard errors, tabulated against $n$.}
\begin{tabular}{c c c c }
\hline
Model&$n=100$&$n=250$&$n=500$\\ [0.5ex] 
\hline
MLTA&$0.48 \,(0.03)$&$0.54 \,(0.02)$&$0.56 \,(0.005)$\\
MCLT&$0.65 \,(0.008)$&$0.54 \,(0.006)$&$0.60 \,(0.004)$\\  [1ex]
\hline
\end{tabular}
\label{table:4.3}
\end{table} 

\begin{figure}[H]
        \centering
        \begin{subfigure}[b]{0.44\textwidth}
                \includegraphics[width=\textwidth]{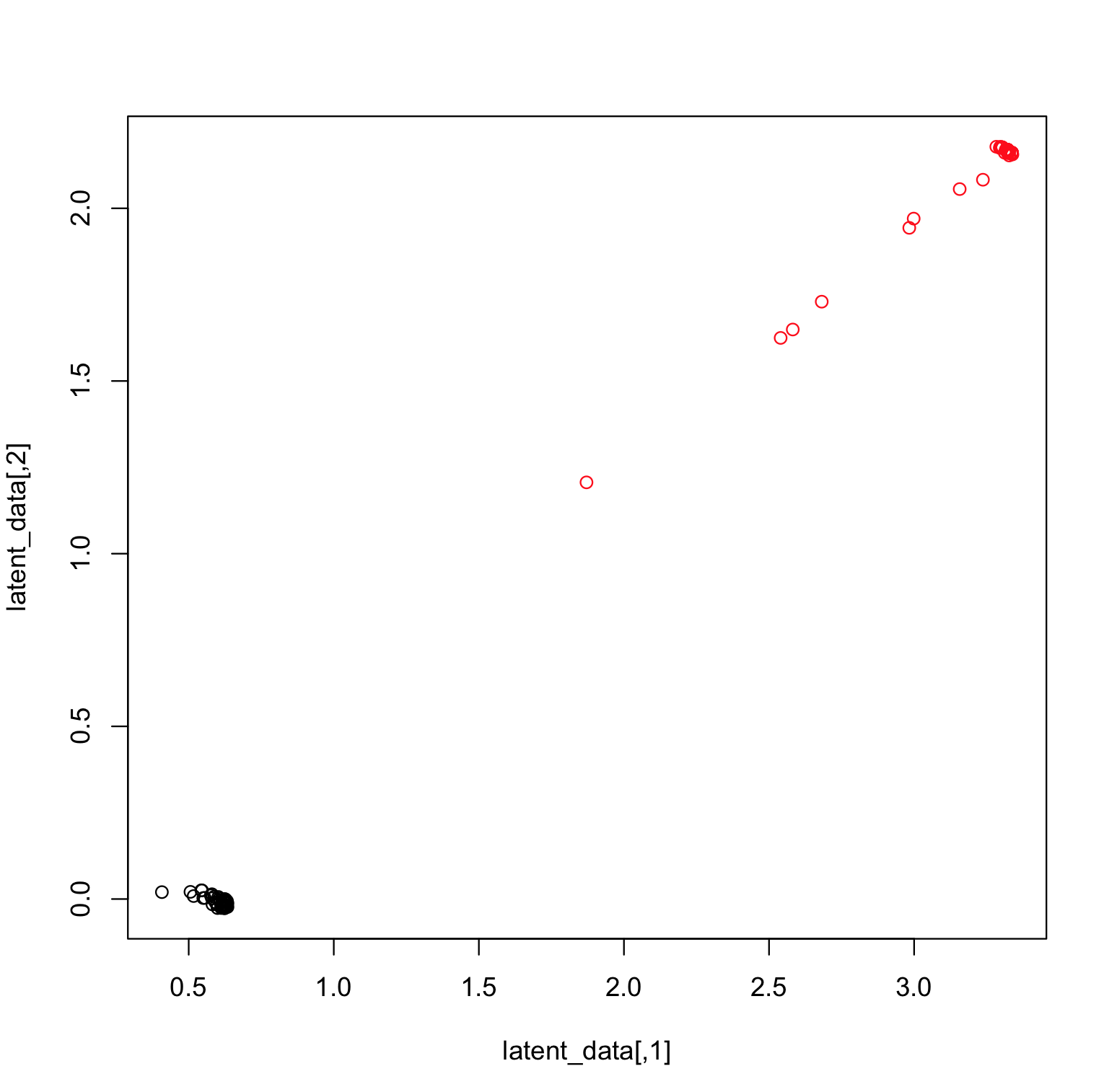}
                \caption{$n=100$}
        \end{subfigure}%
         \qquad
        \begin{subfigure}[b]{0.44\textwidth}
                \includegraphics[width=\textwidth]{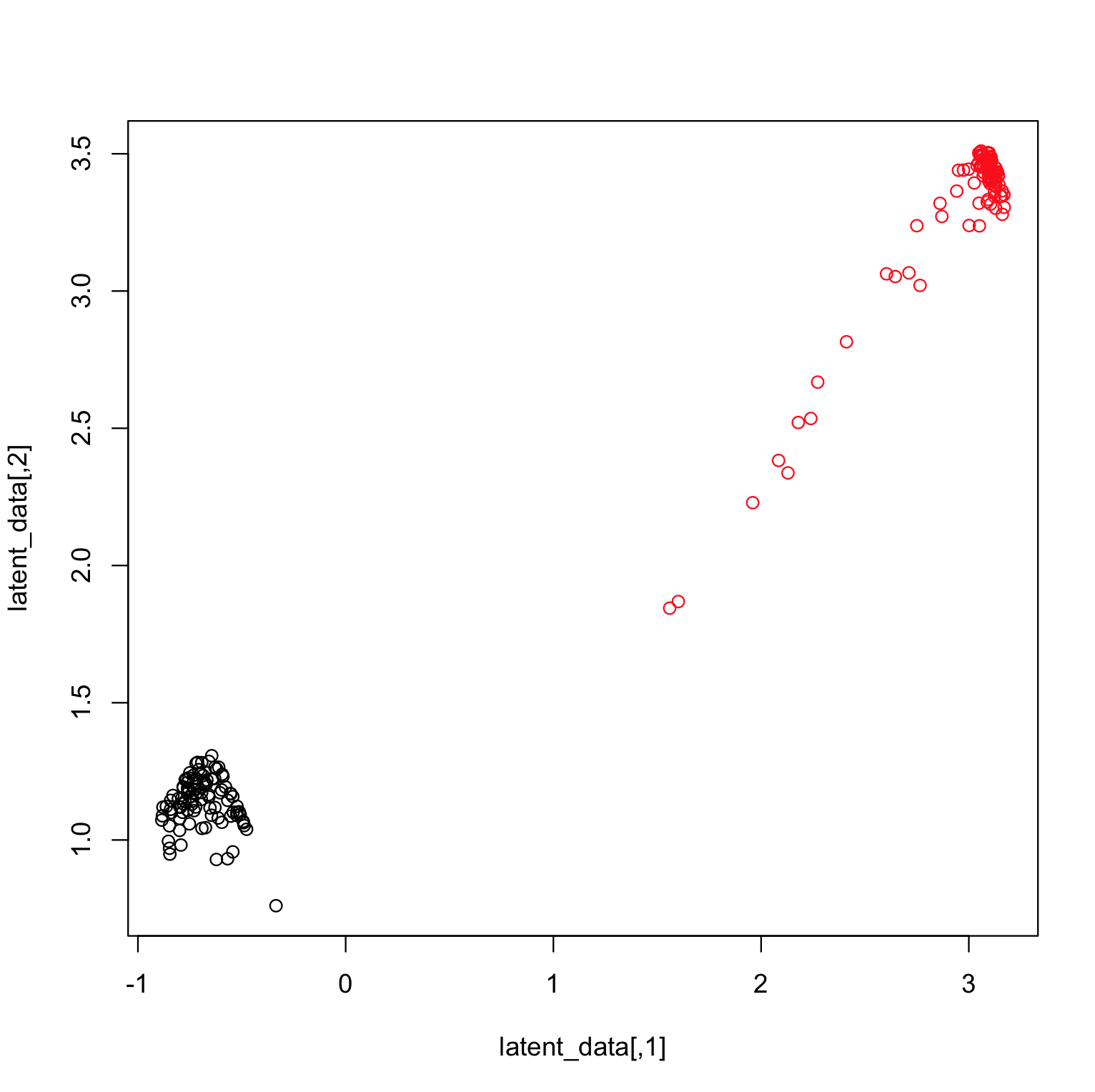}
                \caption{$n=250$}
      
        \end{subfigure}
        ~ 
        \begin{subfigure}[b]{0.44\textwidth}
                \includegraphics[width=\textwidth]{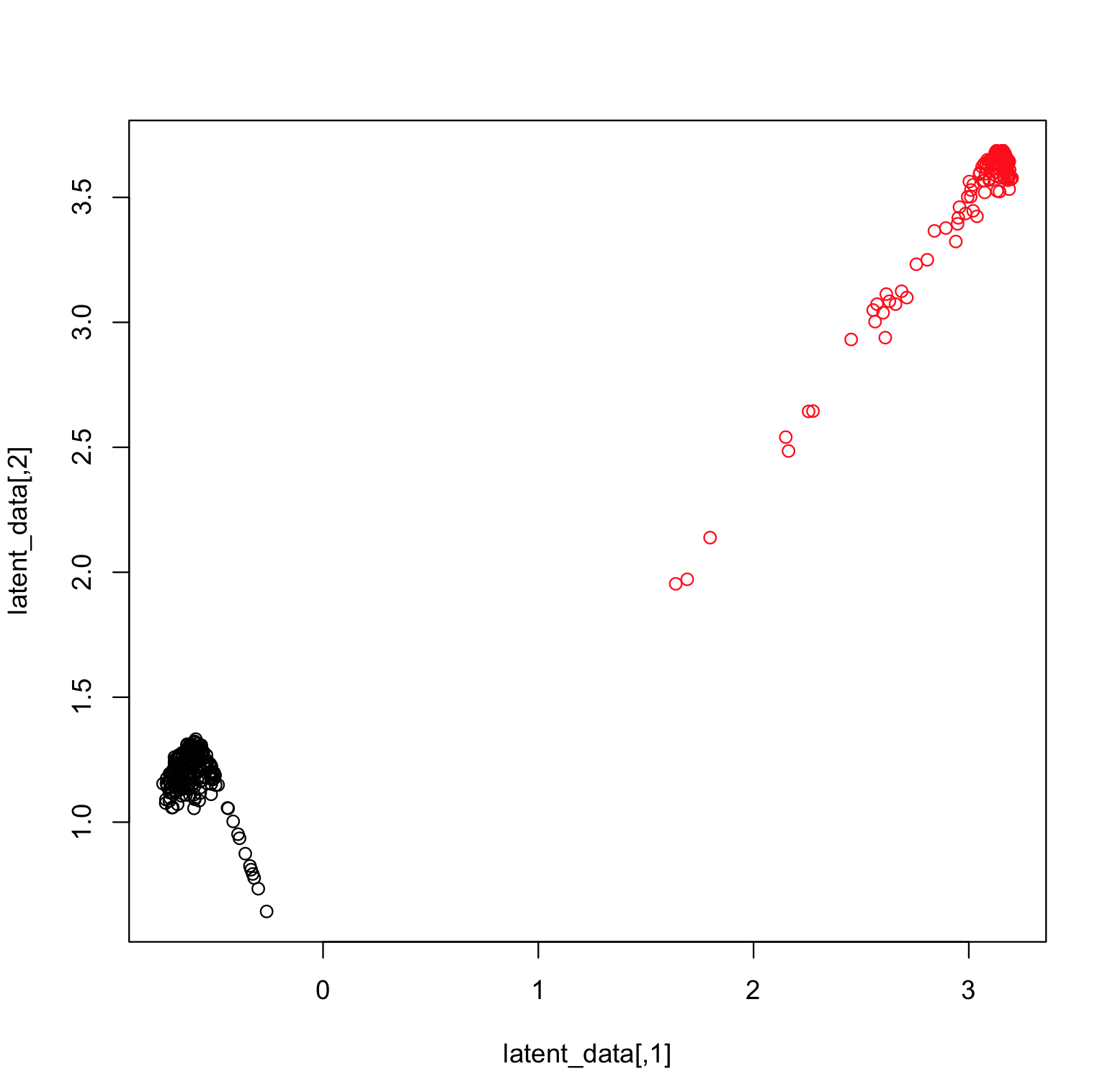}
                \caption{$n=500$}
           
        \end{subfigure}
        \caption{Plots of the estimated posterior mean for different $n$.}\label{fig:4.1}
\end{figure}

We have also given a plot of the estimated posterior mean for each sample size (Figure~\ref{fig:4.1}). These projections are not applicable in MLTA approach as, in its formulation, the latent variables have no cluster-specific discriminatory features.

\section{Application}\label{sec:Application}

\subsection {U.S. Congressional Voting} \label{sec:Voting}
A U.S.\ congressional voting data set \citep{Bache_2013} has been widely used in the literature \citep[e.g.,][]{Gunopulos_2002, Murphy_2013}. This data set includes votes of 435 U.S. House of Representatives congressmen on on sixteen key issues in 1984 with three different type of votes: yes, no, or undecided. The voter's party is labeled as a Democrat or a Republican. The issues voted on are listed in Table~\ref{table:add}.
\begin{table}[htbp]
\centering
 \caption{The issues that were voted on in the U.S.\ congressional voting data.}
  \label{table:add}%
\begin{tabular*}{1\textwidth}{@{\extracolsep{\fill}}rlrl}
\hline
     Item &Issue & Item  & Issue\\
\hline
    1  &Handicapped Infants &  9 &MX Missile\\
   2  & Water Project Cost-Sharing & 10 & Immigration \\
 3  & Adoption of the Budget Resolution &11& Synfuels Corporation Cutback\\
    4& Physician Fee Freeze& 12 & Education Spending\\
    5  & El Salvador Aid  &13 & Superfund Right to Sue\\
  6 & Religious Groups in Schools  &14 &Crime\\
    7  &Anti-Satellite Test Ban &15 & Duty- Free Exports\\
    8 & Aid to Nicaraguan `Contras' & 16 & Export Administration Act/South Africa\\
 \hline
    \end{tabular*}
\end{table}

\begin{figure}[h!]
   \centering
    \includegraphics[width=1\textwidth]{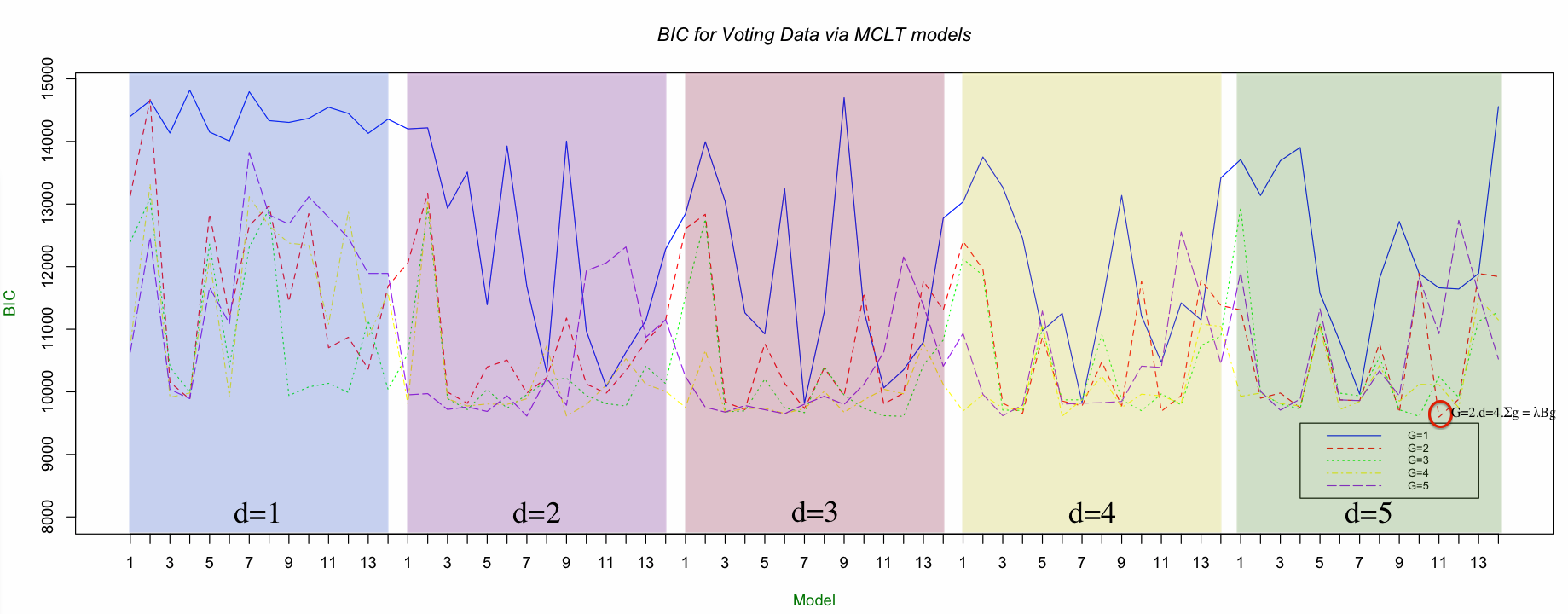}
 \caption{BIC values for all 70 different models fitted to the U.S.\ Congressional voting data for $G=1,G=2\ldots, G=5$. The order of the models in each dimension ($x$ axis) is as same as in Table~\ref{table:3.2}.}
\label{figure5.1}
\end{figure}

We code each question in two binary variables A and B: the responses for the A variables are coded as $1=$ yes/no and $0=$ undecided; and B variables are $1=$ yes, $0=$ no/undecided. 
The fourteen MCLT models 
were fitted to these data for $d=1,2,\ldots, 5$ and $G=1,2,\ldots, 5$. The minimum BIC (Figure~\ref{figure5.1}) occurs at the 2-group, 5-dimensional latent trait model and $\boldsymbol{\Sigma}_g=\lambda \boldsymbol{B}_g$, which is considered as the ``best'' model. The 
the BIC value is $9597$.

\subsubsection{A Comparison of approaches: MLTA vs. MCLT}
The key statistics on the best models for MLTA, PMLTA, and MCLT are shown in Table~\ref{table:5.1}. It can be seen that the highest ARI value ($0.64$) is obtained using the MCLT model. Moreover, the MCLT model gives us fewer groups compared to other approaches. 
\begin{table}[ht]
\centering
\caption{Presents a comparison of 3 different approaches.}
\begin{tabular*}{1\textwidth}{@{\extracolsep{\fill}}c l c c c c c}
\hline
&Model&$G$&$D$&BIC&$\boldsymbol{\Sigma}_g$&ARI\\ [0.5ex] 
\hline
1&MLTA&$3$&$1$&9812&n/a&$0.42$\\
2&MLTA Parsimonious&$4$&$2$&$9681$&n/a&$0.47$\\
3&MCLT&$2$&$5$&9597&EVI&{\bfseries 0.64} \\  [1ex]
\hline
\end{tabular*}
\label{table:5.1}
\end{table}

\subsubsection{Analysis of the Selected MCLT Model}
 The classification table of the group membership with party membership is presented in Table~\ref{table:5.2}.  According to our model selection criteria, BIC=$9597$ is the minimum BIC with the highest ARI value ($0.64$) and, therefore, a 2-components and 5-dimensional latent trait model is selected. In comparison with the true party membership, there are only 42 misclassified Congressmen (i.e., 90.3\% accuracy) with the ``best'' model. Group~1 consists mainly of Republican congressman, and Group~2 consists mainly of Democratic congressman. Table~\ref{table:5.3} shows the median probability $\pi_{mg}(0)$ for each of the groups. The probabilities of a positive response for the A variables (yes/no vs.\ undecided) for the median individuals in all groups are always high with only one exception in Group 2, for variable number 16, where $\pi_{16\,2}(0)=0.70$. Thus, the majority of congressmen voted on most issues, but with a slightly lower voting rate in Group 2 on all issues. Due to the high voting rates, most probabilities given for B variables (yes vs.\ no/undecided) can be interpreted in terms of voting `yes' versus `no'. 
\begin{table}[ht]
\centering
\caption{Cross-tabulation of party and predicted classification for our chosen model (EVI, $G=2$, $d=5$) for the U.S.\ Congressional Voting Data.}
\begin{tabular*}{1\textwidth}{@{\extracolsep{\fill}}l c c c  }
\hline
&1&2 \\ [0.5ex] 
\hline
Republican& 156 &12\\
Democrat&30& 237\\  [1ex]
\hline
\end{tabular*}
\label{table:5.2}
\end{table}

It can be observed that the responses for the median individual in Group 1 are opposite to the ones given by the median individual in Group 2 for most issues. The Republican group (Group 1) tend to give positive responses for the variables 4B, 5B, 6B, 12B, 13B, and 14B. These variables are concerned with the physician fee freeze, El Salvador aid, religious groups in schools, education spending, the superfund right to sue, and crime. The democrat group tend to give positive responses for variables 3B, 7B, 8B, and 9B. These variables are concerned with the adoption of the budget resolutions, the anti-satellite test ban, aid to the Nicaraguan `Contras', and the MX Missile. 
\begin{table}[htbp]
  \centering
  \caption{Probabilities that the median individual in Group $g$ has a positive response for each of 16 votes in the U.S.\ Congressional voting data.}    
 \begin{tabular*}{1\textwidth}{@{\extracolsep{\fill}}crrr|crr}
\hline
  Y/N vs.\ Undecided   & G1    & G2    &    & Y vs.\ N/Undecided      & G1    & G2 \\
\hline
    1A    &      0.99  &      0.96  &       & 1B    &      0.19  &      0.61  \\
    2A    &      0.91  &      0.89  &       & 2B    &      0.50  &      0.41  \\
    3A    &      0.98  &      0.97  &       & 3B    &      0.15  &      0.91  \\
    4A    &      0.99  &      0.96  &       & 4B    &      0.90  &      0.05  \\
    5A    &      0.99  &      0.95  &       & 5B    &      0.98  &      0.10  \\
    6A    &      0.99  &      0.96  &       & 6B    &      0.94  &      0.38  \\
    7A    &      0.99  &      0.97  &       & 7B    &      0.16  &      0.86  \\
    8A    &      0.97  &      0.97  &       & 8B    &      0.08  &      0.93  \\
    9A    &      0.99  &      0.93  &       & 9B    &      0.08  &      0.79  \\
    10A   &      0.99  &      0.97  &       & 10B   &      0.51  &      0.49  \\
    11A   &      0.97  &      0.95  &       & 11B   &      0.21  &      0.43  \\
    12A   &      0.95  &      0.92  &       & 12B   &      0.82  &      0.08  \\
    13A   &      0.96  &      0.93  &       & 13B   &      0.87  &      0.18  \\
    14A   &      0.98  &      0.95  &       & 14B   &      0.97  &      0.27  \\
    15A   &      0.95  &      0.93  &       & 15B   &      0.08  &      0.64  \\
    16A   &      0.90  &      0.70  &       & 16B   &      0.57  &      0.69  \\
\hline
    \end{tabular*}%
  \label{table:5.3}%
\end{table}

\begin{figure}[htb]
   \centering
    \includegraphics[width=0.7\textwidth]{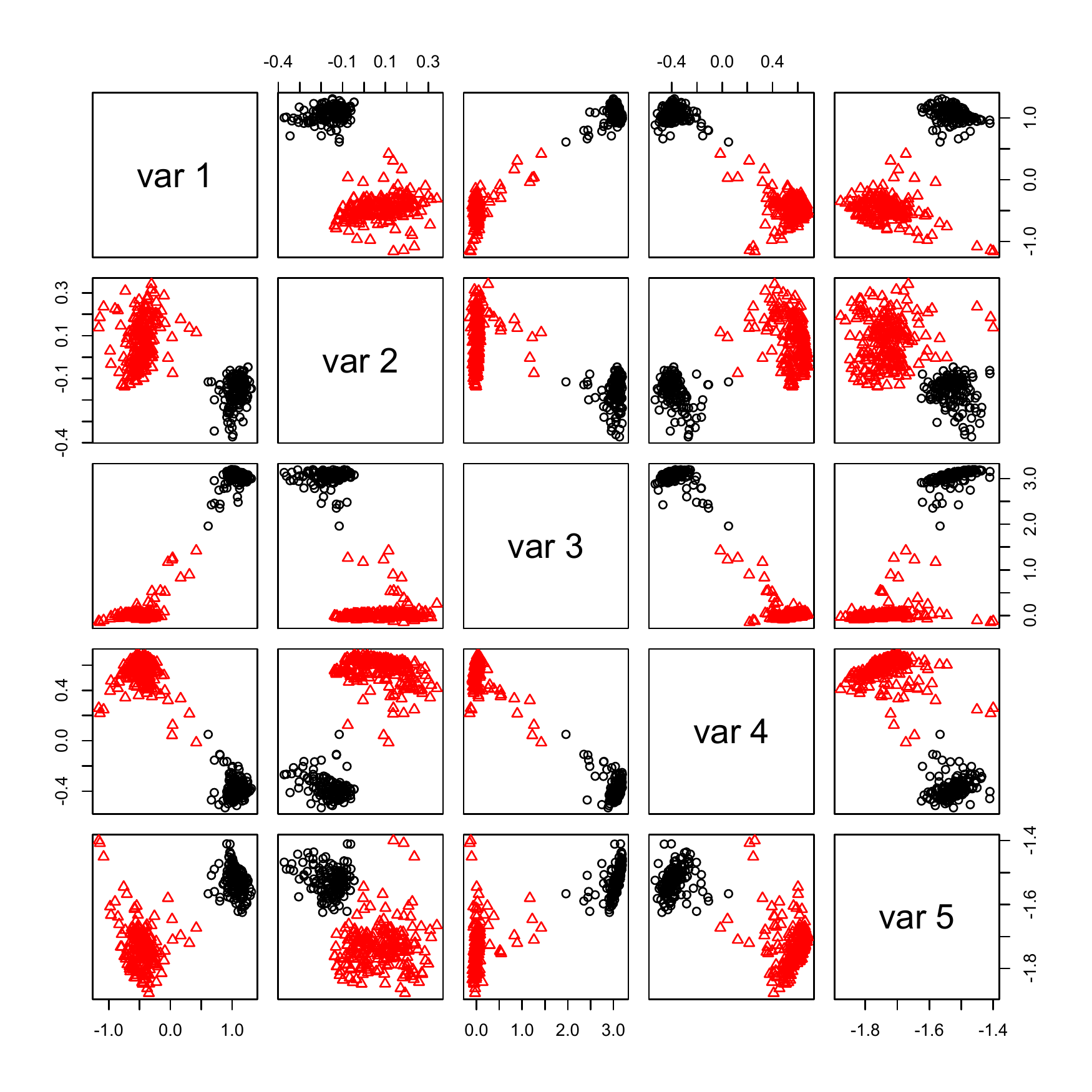}
 \caption[A projection of estimated posterior mean of the best model via MCLT approach with group labels]{Demonstrates the projection of the estimated posterior mean for the \textbf{selected MCLT model} via with group labels.}
\label{figure5.2}
\end{figure}

We have give a plot of the estimated posterior mean of the best MCLT model with group labels (Figure~\ref{figure5.2}). The two groups are well separated, which can be expected because the error rate of the selected model is quite low ($0.093$).

\subsection{Orange Juice Data}\label{sec:OJ}
A data set describing the sensory properties of orange juice is chosen to illustrate the MCLT model with block effect. The data set contains ten commercially available orange juice (OJ) products. One hundred and twenty consumers were recruited, and the tests were conducted over two weeks in a total of four sessions. The choices within the check-all-that-apply (CATA) questions were presented in alphabetical order during week~1 and in Williams design order \citep{Williams_1983} during week~2. In both cases, the attributes were not presented according to sensory modality (appearance, flavour and texture), but in alphabetical order. Therefore, each individual has been accessed 20 times and treated as a block. 
To the end, there are $2400$ observations, of which $100$ are missing. We adopt 40 attributes: 4 in appearance, 27 in flavour, 8 in texture and an indicator for missing observations (Table~\ref{table:5.7}). The study was designed, organized and administered using $\text{Compusense}^{\tiny{\textregistered}}$ at-hand (Compusense Inc., Guelph, ON, Canada).

\begin{table}[H]
\centering
 \caption{Attributes for the orange juice data.}
  \label{table:5.7}%
\begin{tabular*}{1\textwidth}{@{\extracolsep{\fill}}rlrl}
\hline
    Attribute  & Attribute Name& Attribute  & Attribute Name\\
\hline
    A\_1  & A\_Cloudy/Turbid &   F\_18 & F\_Other Citrus Flavor\\
    A\_2  & A\_Orange in Color & F\_19 & F\_Oxidized Flavor \\
    A\_3  & A\_Translucent&F\_20 & F\_Papery/Cardboard Flavor\\
    A\_4  & A\_Yellow & F\_21 & F\_Plastic Flavor\\
    F\_1  & F\_Artificial Flavor  &F\_22 & F\_Processed Flavor\\
    F\_2  & F\_Bitter Taste  &F\_23 & F\_Refreshing Flavor\\
    F\_3  & F\_Cheap Taste &F\_24 & F\_Rotten/Overripe Orange Flavor\\
    F\_4  & F\_Earthy Flavor & F\_25 & F\_Shelf Stable Flavor\\
    F\_5  & F\_Expensive Flavor & F\_26& F\_Strong Flavor\\
    F\_6  & F\_Fresh Orange Flavor  & F\_27 & F\_Weak/Watery Flavor \\
    F\_7  & F\_Fresh Squeezed Flavor  &T\_1  & T\_Astringent/Mouth Drying\\
    F\_8  & F\_From Concentrate Flavor  &   T\_2  & T\_Chunky\\
    F\_9  & F\_Green/Unripe Orange Flavor &T\_3  & T\_Grainy/Chalky\\
    F\_10 & F\_High Acidic/Sour/Tart Taste& T\_4  & T\_Has a Mouthcoat  \\
    F\_11 & F\_High Sweet Taste& T\_5  & T\_Pulpy \\
    F\_12 & F\_Lemon Flavor& T\_6  & T\_Smooth \\
    F\_13 & F\_Low Sweet Taste &T\_7  & T\_Thick\\
    F\_14 & F\_Low Acidic/Sour/Tart Taste &T\_8  & T\_Thin\\
    F\_15 & F\_Natural Flavor \\
    F\_16 & F\_Not From Concentrate Flavor \\
    F\_17 & F\_Organic Flavor \\
   \hline
    \end{tabular*}
\end{table}

We fit MCLT models with block effect to these data for  $d=1,2,\ldots,6$ and $G=1,2, \ldots, 8$. 
The minimum BIC ($72538$) occurs at the 7-group, 4-dimensional latent trait model and $\boldsymbol{\Sigma}_g=\lambda_g\boldsymbol{B}$, which is considered as the ``best'' model (Table~\ref{tab:VEI}).
\begin{table}[htbp]
  \centering
  \caption[BIC for Model VEI]{BIC values for model VEI ($\boldsymbol \Sigma_g=\lambda_g \boldsymbol B$) are listed. }
    \begin{tabular}{crrrrrrrr}
\hline
     Dim/Group& 1     & 2     & 3     & 4     & 5     & 6     & 7     & 8 \\
\hline
    1     & 86709 & 85557 & 83375 & 84605 & 86624 & 81715 & 82404 & 78416 \\
    2     & 83358 & 84556 & 82002 & 80160 & 81757 & 81577 & 76281 & 75280 \\
    3     & 81984 & 80535 & 80083 & 80248 & 80614 & 78168 & 75646 & 75014 \\
    4     & 83914 & 80576 & 79178 & 82180 & 77992 & 76571 & \bf 72538 & 78414 \\
    5     & 84556 & 81765 & 80160 & 84350 & 77039 & 77794 & 75009 & 78800 \\
    6     & 85427 & 83044 & 83703 & 85389 & 75376 & 77986 & 75443 & 80535 \\
\hline
    \end{tabular}%
  \label{tab:VEI}%
\end{table}%

\subsubsection{Analysis of the Selected MCLT Model}
Instead of treating each observation independently, we treat each individual as a block, where each block consists of 20 observations. The classification table of the group membership with product label is presented in Table~\ref{table:5.4}. Group 1 consists mainly of products 4, 6, 7, 10; Group 2 consists mainly products 1, 2, 3, 5, 8; Group 3 consists mainly of the missing observations; Group 4 is a small group consists mainly products 1, 2, 5, 9; Group 5 has only four observations; Group 6 is another small group consists mainly products 4, 7, 10; and Group 7 consists mainly of products 2, 3, 5, 8, 9.  
\begin{table}[H]
  \centering
  \caption{Cross-tabulation of predicted classifications versus product label for the best MCLT model (VEI, $G=7$, $d=4$) applied to the orange juice data.}
\begin{tabular*}{1\textwidth}{@{\extracolsep{\fill}}lrrrrrrr}
\hline
        & $G=1$   & $G=2$   & $G=3$   & $G=4$   & $G=5$   & $G=6$   & $G=7$\\
\hline    
Missing & 0     & 0     & {\bfseries 100}   & 0     & 0     & 0     & 0 \\
    P1    & 48    & {\bfseries 103}   & 42    & 11    & 1     & 0     & 23 \\
    P2    & 25    & {\bfseries 115}   & 21    & 14    & 0     & 1     & 55 \\
    P3    & 21    & {\bfseries 125}   & 21    & 9     & 0     & 4     & 49 \\
    P4    & \bfseries 133   & 48    & 35    & 2     & 0     & 13    & 1 \\
    P5    & 16    & {\bfseries117}   & 27    & 19    & 0     & 0     & 54 \\
    P6    & \bf 170   & 28    & 22    & 0     & 1     & 6     & 5 \\
    P7    & \bf 162   & 35    & 19    & 1     & 0     & 10    & 2 \\
    P8    & 12    & \bf 138   & 20    & 7     & 1     & 0     & 53 \\
    P9    & 39    & 79    & 27    & 13    & 1     & 1     & 68 \\
    P10   & \bf 121   & 67    & 21    & 1     & 0     & 15    & 2 \\
    Average Overall Impression    & 5.34  & 7.5   & 6.11  & 5.97  & 5.75  & 7.46  & 4.91 \\
 \hline    
  \end{tabular*}
  \label{table:5.4}
\end{table}

The groups found in this analysis have similar structures to the ones found using MLTA. By adding the blocking latent variable, we can separate products more accurately. From Table~\ref{table:5.6} it can be seen that Group~1 consists mainly of products that appear yellow, taste artificial, and have thin texture.  The average overall impression score of Group~1 is 5.3/10, which is relatively low among all groups. 
In contrast, Group~2 consists mainly of products that are orange in colour, fresh in flavour, and pulpy in texture. The average overall impression score of Group~2 is 7.5/10 which is the highest among all groups. 
Group~7 is characterized by products are thick, taste bitter, and look cloudy to consumers. The average overall impression score of Group~7 is 4.9/10 which is the lowest among all groups. 
Group~6 is a small group consists of products that are thin but smooth in texture. Group~4 is another small group consists of products have pulpy texture. 
All missing observations fall into Group~3, and all other observations therein have low probability of positive responses for all attributes (cf.\ Table~\ref{table:5.6}). 

Because there are a large number of response patterns with a very small number of observations (of all $2,241$ observed response patterns, only $50$ contain more than one count and only $7$ contain more than two counts), the Pearson's $\chi^2$ test is not applicable. We calculate the overall number of selections (counts) for each attribute across all products to check the goodness of fit. Table~\ref{table:5.5} shows the observed counts and the expected counts for counts over $500$. The table shows that there is a close match between the observed and expected frequencies for most attributes under this model. 
\begin{table}[!htb]
  \centering
  \caption{Observed and expected counts for attributes with 500 or more obsered counts.}
   \begin{tabular*}{1\textwidth}{@{\extracolsep{\fill}}lrrr}
  \hline
 &Attribute         & Observed Counts & Expected Counts \\
\hline
    \multirow{3}{*}{Apperance}&A\_1  & 777   & 653 \\
    &A\_2  & 1364  & 1583 \\
    &A\_4  & 924   & 871 \\
\hline
   \multirow{11}{*}{Flavour} &F\_1  & 539   & 557 \\
    &F\_2  & 529   & 221 \\
    &F\_6  & 757   & 598 \\
    &F\_7  & 595   & 597 \\
    &F\_8  & 623   & 446 \\
    &F\_10 & 563   & 135 \\
    &F\_13 & 614   & 114 \\
    &F\_15 & 586   & 598 \\
    &F\_22 & 512   & 333 \\
    &F\_23 & 608   & 598 \\
    &F\_26 & 717   & 683 \\
\hline
  \multirow{4}{*}{Texture}  &T\_5  & 1128  & 1269 \\
    &T\_6  & 895   & 867 \\
    &T\_7  & 641   & 926 \\
    &T\_8  & 667   & 690 \\
    \hline
    \end{tabular*}%
  \label{table:5.5}%
\end{table}

\begin{figure}[!htb]
   \centering
    \includegraphics[width=0.7\textwidth]{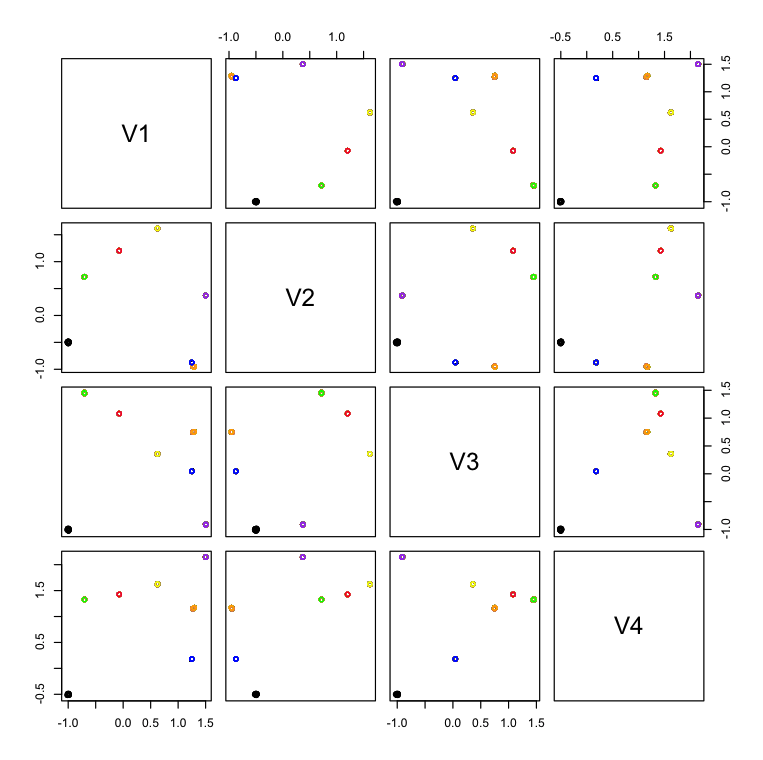}
 \caption[A projection of estimated posterior mean of the best model via MCLT approach with group labels]{Projects the estimated posterior mean of the best model via MCLT approach with group labels.}
\label{figure5.3}
\end{figure}

We have also given the plot of the estimated posterior mean in selected model via MCLT with group labels (Figure~\ref{figure5.3}). Despite the fact that the posterior mean of  Groups~3 and~4 are close together in $d_1$ and $d_2$, all groups are well separated in all dimensions.  

\section{Conclusion}\label{sec:conclusion}
The mixture of latent trait models with common slope parameters gives good clustering performance when applied to high-dimensional binary data. The MCLT model with block effect provides a suitable alternative for clustered data. Our variational EM algorithm gives provided an effective and efficient approach to parameter estimation. 

The MCLT model provides a model-based clustering framework for high-dimensional binary data by drawing on ideas from common factor analyzers. The sharing of the slope parameters enables the model to cluster high-dimensional binary data and to provide low-dimensional plots of the clusters so obtained. The latter plots are given in terms of the (estimated) posterior means of the latent variables. These projections are not applicable in the MLTA approach as, in its formulation, the latent variables have no cluster-specific discriminatory features. The MLTA approach does allow a more general representation of the component covariances and places no restrictions on the component means. However, in this paper, we demonstrate that the MCLT model is useful when the dimension $m$ and the number of clusters $G$ is large. In analogy to the famous GPCM family of mixture models of \citet{Celeux_1995}, c.f. Section~\ref{sec:GPCM}, fourteen covariance structures have been implemented to introduce parsimony. We have presented analyses of two data sets to demonstrate the usefulness of this approach. The model parameters are interpretable and provide a characterization of the within-cluster structure. In our applications herein, we used the BIC to choose the number of clusters $G$, the latent variable dimension $d$, and the covariance decomposition. 

In our future work, we wish to investigate other alternatives for repeatedly sampled data. An alternative model for multi-nominal data can be developed using a mixture polytomous logit model.  

\section*{Acknowledgements}
The authors gratefully acknowledge the helpful comments of the anonymous reviews. This work was supported by a grant-in-aid from Compusense Inc., a Collaborative Research and Development grant from the Natural Sciences and Engineering Research Council of Canada, an Early Researcher Award from the Government of Ontario, and the University Research Chair in Computational Statistics. The authors are grateful to Compusense Inc.\ for granting access to the orange juice data.


\appendix
\section{Additional Table} \label{AppendixA}
\footnotesize
{\scriptsize
\begin{longtable}{lcr|lcr|lcr}
\caption{Probabilities that the median individual in Group $G$ has a positive response for each attribute in the orange juice data.}\label{table:5.6}\\
\hline
    Attribute & Group &  $\pi_{mg}(0)$    & Attribute & Group &  $\pi_{mg}(0)$    & Attribute & Group & $\pi_{mg}(0)$ \\
\hline
    A\_1  & 1     &      0.26  & F\_10 & 1     &      0.37  & F\_23 & 1     &      0.05  \\
          & 2     &      0.30    &       & 2     &      0.10  &       & 2     &      0.61 \\
          & 3     &      0.27  &       & 3     &      0.16  &       & 3     &      0.06  \\
          & 4     &      0.33    &       & 4     &      0.06  &       & 4     &      0.19  \\
          & 5     &      0.39    &       & 5     &      0.25  &       & 5     &      0.18  \\
          & 6     &      0.10    &       & 6     &      0.10  &       & 6     &      0.54  \\
          & 7     &      0.62    &       & 7     &      0.49  &       & 7     &      0.02  \\
    A\_2  & 1     &      0.43    & F\_11 & 1     &      0.13  & F\_24 & 1     &      0.08  \\
          & 2     &       0.72    &       & 2     &      0.22  &       & 2     &      0.03  \\
          & 3     &      0.40  &       & 3     &      0.09  &       & 3     &      0.03  \\
          & 4     &      0.08   &       & 4     &      0.21  &       & 4     &      0.11  \\
          & 5     &      0.67    &       & 5     &      0.18  &       & 5     &      0.10  \\
          & 6     &      0.19   &       & 6     &      0.23  &       & 6     &      0.02  \\
          & 7     &      0.90    &       & 7     &      0.09  &       & 7     &      0.19  \\
    A\_3  & 1     &      0.14   & F\_12 & 1     &      0.12  & F\_25 & 1     &      0.07  \\
          & 2     &      0.07  &       & 2     &      0.05  &       & 2     &      0.04  \\
          & 3     &      0.04    &       & 3     &      0.03  &       & 3     &      0.02  \\
          & 4     &      0.06    &       & 4     &      0.05  &       & 4     &      0.04  \\
          & 5     &      0.09    &       & 5     &      0.10  &       & 5     &      0.06  \\
          & 6     &      0.22  &       & 6     &      0.06  &       & 6     &      0.08  \\
          & 7     &      0.03    &       & 7     &      0.10  &       & 7     &      0.03  \\
    A\_4  & 1     &      0.50   & F\_13 & 1     &      0.28  & F\_26 & 1     &      0.26  \\
          & 2     &      0.27   &       & 2     &      0.22  &       & 2     &      0.34  \\
          & 3     &      0.48    &       & 3     &      0.23  &       & 3     &      0.18  \\
          & 4     &      0.78  &       & 4     &      0.30  &       & 4     &      0.07  \\
          & 5     &      0.27    &       & 5     &      0.29  &       & 5     &      0.35  \\
          & 6     &      0.81    &       & 6     &      0.20  &       & 6     &      0.15  \\
          & 7     &      0.07   &       & 7     &      0.32  &       & 7     &      0.51  \\
    F\_1  & 1     &      0.50  & F\_14 & 1     &      0.18  & F\_27 & 1     &      0.28  \\
          & 2     &      0.04  &       & 2     &      0.23  &       & 2     &      0.03  \\
          & 3     &      0.16  &       & 3     &      0.17  &       & 3     &      0.09  \\
          & 4     &      0.15  &       & 4     &      0.47  &       & 4     &      0.21  \\
          & 5     &      0.18  &       & 5     &      0.22  &       & 5     &      0.12  \\
          & 6     &      0.15  &       & 6     &      0.26  &       & 6     &      0.12  \\
          & 7     &      0.30  &       & 7     &      0.13  &       & 7     &      0.12  \\
    F\_2  & 1     &      0.39  & F\_15 & 1     &      0.03  & T\_1  & 1     &      0.27  \\
          & 2     &      0.03  &       & 2     &      0.60  &       & 2     &      0.05  \\
          & 3     &      0.12  &       & 3     &      0.05  &       & 3     &      0.08  \\
          & 4     &      0.07  &       & 4     &      0.18  &       & 4     &      0.05  \\
          & 5     &      0.20  &       & 5     &      0.20  &       & 5     &      0.14  \\
          & 6     &      0.04  &       & 6     &      0.25  &       & 6     &      0.12  \\
          & 7     &      0.57  &       & 7     &      0.04  &       & 7     &      0.18  \\
    F\_3  & 1     &      0.42  & F\_16 & 1     &      0.06  & T\_2  & 1     &      0.03  \\
          & 2     &      0.02  &       & 2     &      0.25  &       & 2     &      0.10  \\
          & 3     &      0.10  &       & 3     &      0.04  &       & 3     &      0.03  \\
          & 4     &      0.12  &       & 4     &      0.09  &       & 4     &      0.21  \\
          & 5     &      0.13  &       & 5     &      0.15  &       & 5     &      0.16  \\
          & 6     &      0.11  &       & 6     &      0.12  &       & 6     &      0.01  \\
          & 7     &      0.20  &       & 7     &      0.07  &       & 7     &      0.32  \\
    F\_4  & 1     &      0.06  & F\_17 & 1     &      0.03  & T\_3  & 1     &      0.09  \\
          & 2     &      0.12  &       & 2     &      0.12  &       & 2     &      0.04  \\
          & 3     &      0.03  &       & 3     &      0.01  &       & 3     &      0.03  \\
          & 4     &      0.07  &       & 4     &      0.07  &       & 4     &      0.10  \\
          & 5     &      0.11  &       & 5     &      0.09  &       & 5     &      0.10  \\
          & 6     &      0.06  &       & 6     &      0.05  &       & 6     &      0.02  \\
          & 7     &      0.08  &       & 7     &      0.04  &       & 7     &      0.14  \\
    F\_5  & 1     &      0.02  & F\_18 & 1     &      0.23  & T\_4  & 1     &      0.22  \\
          & 2     &      0.24  &       & 2     &      0.08  &       & 2     &      0.10  \\
          & 3     &      0.01  &       & 3     &      0.10  &       & 3     &      0.08  \\
          & 4     &      0.05  &       & 4     &      0.10  &       & 4     &      0.05  \\
          & 5     &      0.09  &       & 5     &      0.17  &       & 5     &      0.18  \\
          & 6     &      0.09  &       & 6     &      0.10  &       & 6     &      0.10  \\
          & 7     &      0.02  &       & 7     &      0.22  &       & 7     &      0.24  \\
    F\_6  & 1     &      0.05  & F\_19 & 1     &      0.06  & T\_5  & 1     &      0.10  \\
          & 2     &      0.75  &       & 2     &      0.02  &       & 2     &      0.68  \\
          & 3     &      0.09  &       & 3     &      0.01  &       & 3     &      0.36  \\
          & 4     &      0.22  &       & 4     &      0.05  &       & 4     &      0.71  \\

          & 5     &      0.27  &       & 5     &      0.06  &       & 5     &      0.61  \\
          & 6     &      0.44  &       & 6     &      0.02  &       & 6     &      0.02  \\
          & 7     &      0.06  &       & 7     &      0.07  &       & 7     &      0.89  \\
    F\_7  & 1     &      0.02  & F\_20 & 1     &   0.07  & T\_6  & 1     &   0.50  \\
          & 2     &      0.61  &       & 2     &      0.02  &      & 2     &      0.44  \\
          & 3     &      0.05  &       & 3     &      0.02  &       & 3     &      0.27  \\
          & 4     &      0.22  &       & 4     &      0.06  &       & 4     &      0.17  \\
          & 5     &      0.23  &       & 5     &      0.06  &       & 5     &      0.25  \\
          & 6     &      0.12  &       & 6     &      0.02  &       & 6     &      0.93  \\
          & 7     &      0.08  &       & 7     &      0.07  &       & 7     &      0.03  \\
    F\_8  & 1     &      0.43  & F\_21 & 1     &      0.07  & T\_7  & 1     &      0.06  \\
          & 2     &      0.11  &       & 2     &      0.02  &       & 2     &      0.35  \\
          & 3     &      0.25  &       & 3     &      0.01  &       & 3     &      0.14  \\
          & 4     &      0.34  &       & 4     &      0.04  &       & 4     &      0.49  \\
          & 5     &      0.26  &       & 5     &      0.06  &       & 5     &      0.37  \\
          & 6     &      0.24  &       & 6     &      0.04  &       & 6     &      0.02  \\
          & 7     &      0.30  &       & 7     &      0.04  &       & 7     &      0.65  \\
    F\_9  & 1     &      0.15  & F\_22 & 1     &      0.43  & T\_8  & 1     &      0.58  \\
          & 2     &      0.03  &       & 2     &      0.04  &       & 2     &      0.13  \\
          & 3     &      0.04  &       & 3     &      0.17  &       & 3     &      0.21  \\
          & 4     &      0.07  &       & 4     &      0.20  &       & 4     &      0.10  \\
          & 5     &      0.10  &       & 5     &      0.19  &       & 5     &      0.18  \\
          & 6     &      0.04  &       & 6     &      0.14  &       & 6     &      0.79  \\
          & 7     &      0.14  &       & 7     &      0.30  &       & 7     &      0.05  \\
\hline
\end{longtable}}

\end{document}